\documentclass[a4paper,twocolumn]{aastex62}

\usepackage{natbib}

\usepackage{times}

\usepackage{graphicx}	

\renewcommand{\vec}[1]{\mbox{\boldmath$#1$}}
\newcommand{\cpddd}[0]{CPD-28$^\circ$ 2561}
\newcommand{\zetpup}[0]{$\zeta$ Pup}
\newcommand{\tetori}[0]{$\theta^1$ Ori C}
\newcommand{\sigori}[0]{$\sigma$ Ori E}
\newcommand{\msol}[0]{$M_\odot$}
\newcommand{\confin}[0]{$\eta_*$}

\date{\today}

\submitjournal{Astronomische Nachrichten}

\begin{document}

\author{Manfred K\"uker}
\affil{Leibniz-Institut für Astrophysik Potsdam \\
An der Sternwarte 16, 14482 Potsdam, Germany}

\title{Mass loss and magnetospheres of massive stars}


\begin{abstract}
About ten percent of all OB stars show strong, large-scale surface magnetic fields. The interaction of the magnetic field and the wind is believed to be the cause for the X-ray emission shown by these objects. We therefore run numerical simulations in two dimensions for a number the O-type stars and Wolf Rayet stars to study the interaction of the stellar magnetic fields of these stars with their winds. While weak, dipolar  magnetic fields leave  the wind largely unmodified and the field opens up and becomes a split monopole, the interaction between magnetic field and outflowing gas is more complex
as the magnetic field remains closed in some regions and outflowing gas can be trapped. We use the Nirvana MHD code with adaptive mesh refinement  to study this interaction with high numerical resolution to model cases with confinement parameters up to $10^4$.
\end{abstract}

\keywords{stars: magnetic fields, stars: mass loss, stars: massive, magnetohydrodynamics}

\section{Introduction}
Massive stars lose mass through line-driven winds. 
A fraction of about 10\% of these stars also show large-scale surface magnetic fields with field strengths of the order 1 kG \citep{hubrig11, wade16}. The origins of these magnetic fields are unknown. Unlike lower main sequence stars, massive stars have convective cores and radiative envelopes, which requires a different field generation process. As there are thin convective shells embedded in the envelope that are caused by the iron opacity bump \citep{iglesias92}, dynamo action in these convection zones has been proposed \citep{cantiello09}. While such a dynamo could support the observed field strengths, thin shell dynamos do not usually produce large-scale field geometries.  \cite{ruediger95} showed for an $\alpha \, \Omega$ dynamo at the bottom of the solar convection zone that the spatial field structures and cycle time scales are too small to make that model a viable explanation of the solar activity cycle. The alternative scenario of a turbulent dynamo does not produce ordered large-scale fields either, cf.~\cite{brandenburg12}.
 
Massive stars are the sources of X-rays that are believed to be generated by three mechanisms: shocks in the radiation-driven winds originating from these stars \citep{feldmeier97}, collisions of winds in binary systems \citep{stevens92, pittard09}, and the confinement of a wind by the stellar magnetic field \citep{babel97b,babel97a}. The  magnetically confined wind shock (MCWS) model  explains the moderately hard X-ray spectra of magnetic B- and A-type stars with shocks that form in the equatorial plane of the stellar magnetic field as the gas flows from the two hemispheres collide \citep{robrade11}. However, it takes either very strong magnetic fields or low mass loss rates to trap the gas  inside an unperturbed dipole field and force the gas flow to follow the field line. The MCWS model can explain the X-ray spectrum of the O-type star $\theta^1$ Ori C \citep{gagne05} but the larger mass loss rates of O-type stars will generally lead to a significant distortion of the stellar magnetic field even close to the surface and at large radii the wind will dominate over the magnetic field. 

The first theoretical model of the solar wind by \citet{parker58} already assumes that far 
away from the Sun the magnetic field is essentially radial and varies with radius as $1/r^2$. Potential field extrapolations of the magnetic field in the solar corona with a zero potential at 2.5 solar radii confirm this assumption \citep{altschuler69}. The relative strength of gas flow and magnetic field is given by the Alfv\'en Mach number,   $ M_A = \sqrt{(4 \pi \rho v^2)/B_r^2}$ \citep{weber67}, which assumes values less than one close to the star and greater than one far away from the star. The radius where $M_A=1$  is called the Alfv\'en radius and marks the point where the gas flow becomes dominant over the magnetic field.  In the context of massive stars, \cite{udd02} introduce the quantity $\eta = {1}/{M_{\rm A}^2} $. Replacing local quantities with global model properties such as the magnetic field strength on the stellar surface at the equator, $B_{\rm eq},$ the stellar radius, $R_*,$ the mass loss rate, $\dot{M},$ and the asymptotic value of the radial gas velocity, $v_\infty$, leads to the definition of the confinement parameter,
\begin{equation}
   \eta_* = \frac{B_{\rm eq}^2 R_*^2}{\dot{M} v_\infty}.
\end{equation} 
Values greater than one indicate systems where a region with closed field lines exists
while for \mbox{$\eta_*\ll1$} the wind solution is only significantly affected in the equatorial plane.


The MCWS model was developed and tested for OB stars, but, to our knowledge, was not yet applied for the studies of wind confinement in stars with very strong wind, such as Wolf-Rayet (WR) stars. These are evolved massive stars with high mass-loss rates. Their X-ray properties are different from those of O-type stars ~\citep{ignace00, oskinova12}.
Scenarios for the generation of X-rays in the winds of single WR stars include line-driven wind instabilities, MCWS, and corotating interaction regions (CIR). 
Because of the high mass loss rates and smaller radii, the surface field strengths needed to confine WR star winds are even higher than for OB stars, usually of the order $10^4$ G \citep{shenar14}. 
Surface field strengths inferred from spectropolarimetric measurements in the visible part of the wind have so far always been much weaker than that \citep{hubrig16}. Still, magnetic fields can not be ruled out as a cause of X-ray emission. 

The interaction between a dipolar magnetic field rooted in a massive star and the line-driven outflows from that star were first studied by \cite{udd02} 
for an axisymmetric setup using the Zeus MHD code \citep{zeus1,zeus2,zeus3}. Subsequent papers included the effects of rotation \cite{udd08, udd09}, radiative cooling \cite{gagne05}, and 3D geometry \citep{udd13}. All these studies were carried out with the Zeus code. While Zeus is a proven code that has been used in many contexts over a quarter of a century, it lacks some features of more modern codes. Firstly, the Zeus code in its original form is not conservative, which can lead to problems when shocks are involved \citep{falle02, tasker08}. More importantly in the current context, it does not allow the numerical resolution to be adapted to complex geometries or certain features of the solution. 
We therefore present an alternative numerical model of the interaction between radiation-driven winds from massive stars and large-scale stellar magnetic fields using the Nirvana MHD code \citep{ziegler04, ziegler05}. 
%
%
\section{Method}
The Nirvana MHD code  solves the MHD and heat transport equations in conservative form. The primitive variables are the momentum density $\rho \vec{v}$, the magnetic field $\vec{B}$, the mass density $\rho$, and the (total) energy density $e$.
The code solves the equation of motion, 
\begin{eqnarray} 
  \frac{\partial (\rho \vec{v})}{\partial t} + \nabla \cdot \left[ \rho \vec{v} \vec{v}
  + \left(p + \frac{1}{8 \pi}|\vec{B}|^2\right) I - \frac{1}{4 \pi} \vec{B} \vec{B} \right] = \nonumber \\
   \nabla \cdot \tau + \rho \vec{f}_e, + \vec{f}_{\rm cc} \label{motion}
\end{eqnarray}
the induction equation, 
\begin{equation}
  \frac{\partial \vec{B}}{\partial t} - \nabla \times (\vec{v}\times\vec{B} - \eta \nabla \times \vec{B} ) = 0, \label{induction}
\end{equation}
the equation of mass conservation,
\begin{equation}
  \frac{\partial \rho}{\partial t} + \nabla \cdot (\rho \vec{v}) = 0,
\end{equation}
and the equation of energy conservation,
\begin{eqnarray}
   \frac{\partial e}{\partial t} + \nabla \cdot \left[\left(e+p+\frac{1}{8 \pi} |\vec{B}|^2\right)\vec{v} - \frac{1}{4\pi} (\vec{v}\cdot \vec{B}) \vec{B} \right]   \nonumber \\
   = \nabla \cdot \left[ \vec{v} \tau + \frac{\eta}{4\pi} \vec{B} \times(\nabla \times  \vec{B})    - \vec{F}_{\rm cond} \right] \\
   +\rho  \vec{f}_{\rm e} \cdot \vec{v} +\vec{f}_{\rm cc}\cdot\vec{v} + {\cal L}. \nonumber 
    \label{energy}
\end{eqnarray}
In above equations, $\eta$ is the magnetic diffusivity coefficient, $\cal L$ the heat loss function, $\vec{f}_e$ is the sum of the (external) forces acting on the gas, including (stellar) gravity, 
\begin{equation}
  \vec{f}_{\rm cc} = -2 \rho \vec{\Omega} \times \vec{v} - \rho \vec{\Omega} \times (\vec{\Omega} \times \vec{r}) 
\end{equation}
is the sum of the Coriolis and centrifugal forces in the rotating frame of reference.
\begin{equation}
  \tau=\nu (\nabla \vec{v} + (\nabla \vec{v})^{\rm T} - \frac{2}{3} (\nabla \cdot \vec{v}) I) 
\end{equation}
with the kinematic viscosity coefficient $\nu$ is the viscous stress tensor and
\begin{equation}
 \vec{F}_{\rm cond} = \kappa \nabla T 
 \end{equation} 
 with the heat conductivity coefficient $\kappa$ is the conductive heat flux. 
The total energy density is the sum of the thermal, kinetic, and magnetic energy density:
\begin{equation}
  e=\epsilon + \frac{\rho}{2} \vec{v}^2 + \frac{1}{2 \mu} \vec{B}^2.
\end{equation}
For a fully ionised monatomic ideal gas, the equation of state is 
\begin{equation}
   p= \frac{\cal R}{\mu} \rho T
\end{equation} 
with a constant mean molecular weight $\mu$.

The thermal energy density  takes the form
\begin{equation}
  \epsilon = \rho T \frac{\cal R}{\gamma-1}  \label{thermal}
\end{equation}   
with $\gamma=c_p/c_v=5/3$.

The gas in the stellar vicinity is subject to stellar gravity, 
acceleration by electron scattering, and acceleration by line absorption. A prescription for the line force has been derived by \cite{CAK75}, hereafter CAK, for the spherically symmetric case.   Lacking a more accurate formulation for the two-dimensional case, we use the CAK model in the formulation of \cite{udd02}, which includes the corrections by \cite{abbott82} and \cite{pauldrach86} and uses the line distribution normalisation of \cite{gayley95}. 
Gravity and electron scattering are combined into a single force term, so that the sum of the external forces reads 
\begin{equation}
  {f_e} = - \frac{GM(1-\Gamma)}{r^2} + g_{\rm lines},
\end{equation}
where $G$ is Newton's constant of gravity, $M$ the stellar mass, and 
\begin{equation}
  \Gamma= \frac{\kappa_e L}{4 \pi G M c}
\end{equation}
the Eddington parameter with $L$ the stellar luminosity and $\kappa_e$ the electron scattering opacity.
The line force,
\begin{equation}
  g_{\rm lines} = \frac{f}{1-\alpha} \frac{\kappa_e L \bar{Q}}{4 \pi r^2} 
                 \left( \frac{ dv_r/dr}{\rho c \bar{Q}\kappa_e} \right)^\alpha,
                 \label{lineforce}
\end{equation}
depends on the radial velocity component only but can vary with latitude.
The parameter $\bar{Q}$ is related to the $k$ parameter of the original CAK formulation through
\begin{equation}
   k=  \frac{ \bar{Q}^{1-\alpha}  ({v_{\rm th}}/{c})^\alpha }{1-\alpha}.
\end{equation}
The parameters $\alpha$ and $\bar{Q}$ are chosen such that the mass loss rate and terminal wind speed fit the values observed for the object to be modelled. Typical values are $\alpha=0.6$ and $\bar{Q}=500$. 

The Nirvana code uses a Godunov-type volume discretisation scheme that conserves the total energy, which makes it particularly suitable for the treatment of problems where the mass density is low and the gas velocities of the order of the sound speed, i.e.~where compressibility is important and shocks may occur. The code can treat ideal MHD as well as problems with diffusion. It allows the use of cartesian as well as cylindrical and spherical polar coordinates in two and three dimensions. Angular momentum is conserved when polar coordinates are used. 
The code runs on distributed memory parallel computers using the message passing interface (MPI) and allows for adaptive mesh refinement.
In the context of stellar winds we use use the spherical polar coordinates $r$, $\theta$, and $\phi$. In this paper we treat the axisymmetric non-rotating case only but rotation can be added and the model setup can be used in three dimensions with periodic boundary conditions in $\phi$ without modification.

A typical size for the mesh is $N_r=256$ and  $N_\theta=128$ for the basic mesh. The spacing is equidistant in both coordinates. Adaptive mesh refinement brings up the effective mesh size to much higher numbers. We allow for five refinement levels, corresponding to an effective mesh size of $8192\times4096$  grid points. Mesh refinement applies whenever either the mass density, the gas velocity, or the magnetic field changes substantially over short distances, which happens mainly close to the stellar surface and the equatorial plane. For strong magnetic fields, we also find a high mesh resolution near the rotation axis.
%
%
%
%
\section{Results}
\subsection{Model setup}
To verify  our model and for comparison we  choose a setup that is close to that of \cite{udd02}
\subsubsection{Initial conditions}
We start with a spherically-symmetric mass distribution and a dipolar magnetic field. 
The initial velocity is given by the asymptotic velocity law,
\begin{equation}
  v_r=v_\infty   \left(1 - \frac{R}{r} \right)^\beta, \; r>R
  \label{asymptotic}
\end{equation}  
where $v_\infty$ is the terminal wind speed from a run without magnetic field and $\beta$ is a dimensionless parameter of order unity. On the inner boundary, $r=R$, the radial velocity follows from the mass loss rate (see below).
The density profile then follows from mass conservation,
\begin{equation}
   \rho=\frac{\dot{M}}{4 \pi r^2 v_r}, \label{inidens}
\end{equation} 
where $\dot{M}$ is the mass loss rate. In this paper, we assume that the gas is isothermal. 

As the slow stellar rotation is neglected, the magnetic field is the only cause of departure from spherical symmetry and the dipole moment defines the symmetry axis of the system and the coordinate system is aligned with the magnetic field:
\begin{eqnarray}
   B_r      &=& \frac{B_0 \cos \theta \,R^3}{r^3} \\
   B_\theta &=& \frac{B_0 \sin \theta \, R^3}{2 r^3}.
\end{eqnarray}
$B_0$ sets the polar field strength on the stellar surface. 

Figure\ref{eta0_wind} shows the radial velocity and mass density of the initial state as functions of the radius.
\subsubsection{Boundary conditions}
On the inner boudary, we keep the radial component of the magnetic field fixed to its initial values and require that $\partial B_\theta / \partial r = 0$. 
In the momentum equation we require that 
\begin{equation}
  \frac{\partial (r^2 \rho v_r)}{\partial r} = 0   \label{inflow}
\end{equation}
and 
\begin{equation}
  v_\theta = 0 \label{parflow}
\end{equation}
to enforce a gas flow parallel to the magnetic field.  
The density is kept fixed to its initial value $\rho_0$, chosen sufficiently large to ensure subsonic gas flow.
At the outer boundary we use the Nirvana code's built-in outflow boundary conditions. 
As the system is not rotating the azimuthal components of the velocity and magnetic field are zero.
The boundaries in the $\theta$ coordinate are located at $\theta = 0$ and $\theta=\pi$, i.e. on the system's symmetry axis as defined by the dipolar magnetic field. The boundary conditions here are implied by symmetry:
\begin{equation}
  \frac{\partial B_r}{\partial \theta} = B_\theta = 0
\end{equation} 
for the magnetic field,
\begin{equation}
  \frac{\partial (\rho v_r)}{\partial \theta} = (\rho v_\theta) = 0
\end{equation}
for the momentum density and
\begin{equation}
 \frac{\partial \rho}{\partial \theta} = 0
\end{equation} 
 for the mass density.   
\subsection{Applications}   
Our first application is the \zetpup{} model from \cite{udd02}. We then study models for the O star \tetori{}, the Of?p star \cpddd{}, the B2Vp star \sigori{}, and a sample of model stars chosen to cover a range of temperatures, terminal speeds, and mass loss rates typical of Wolf Rayet stars. The model parameters used here are listed in Table \ref{parameters}.  For \zetpup{}, the wind parameters  have been adopted from \cite{udd02}, 
for the remaining stars we have varied them to fit stellar and wind properties from literature. Our model of \tetori{} follows \cite{gagne05} except for the mass of 33.5 \msol found for the primary component by \cite{balega15}. Stellar and wind parameters as well as magnetic field strength for \cpddd{} are from \cite{wade15}, those for \sigori{} from \cite{townsend13}, and those of the the WR stars from \cite{hamann06} and \cite{hubrig16}.  Masses for the WR stars were computed using the mass-luminosity relation for H-free WR stars as discussed in \cite{crowther07}. The values for $B_0$ listed in the table refer to the polar field strength needed for a value of one for the confinement parameter. The total run time is typically chosen to be about an order of magnitude longer than the travel time through the simulation box.
%
\begin{table*}
\caption{Stellar and wind parameters used in this paper. The second column lists the units. The $B_0$ value is for $\eta_*=1$.}
\label{parameters}
\begin{center}
\begin{tabular}{l|ccccccc}
Star & $\zeta$ Pup &$\theta^1$ Ori C & {\cpddd{}}&{\sigori{}} &WR3 & WR6& WR105 \\ \hline
Type & O4I(n)fp& O7Vp C&Of?p& B2Vp &WN3-w & WN4-s &   WN9  \\ 
$M$ ($M_\odot$)  & 46 & 33.5& 61&8.3 &11 & 19 &18.3 \\
$L$ ($10^6 L_\odot$) & $1$ & 0.2& 0.22 &3.57$\times10^{-3}$ &0.38 & 0.4 &   0.5   \\
$T_{\rm eff}$ (kK) & 50 & 39 &35&23&85 & 89 & 32 \\
$R$ $ (R_\odot)$ & 18.6 & 10.6 & 12.9&5.2&2.86 & 2.71 & 21.4  \\
$\dot{M}$ $ (10^{-6} M_\odot$/a) &2.6  & 0.4&1.0&1.5$\times10^{-4}$  & 5 & 50 & 16  \\
$v_\infty$ (km/s) & 2300& 2500& 2400 &  1800 &2200 &1950 & 700 \\
$\bar{Q}$  & 500 & 850& 200& 128   &2000 &  400& 270  \\
$\alpha $  & 0.6 & 0.51&0.43&  0.4   &0.6 & 0.39& 0.45  \\
$B_0 ({\rm kG} )$& 0.3&0.265&0.122 & 9.5   &   2.63 & 8.4 & 0.35  \\
\end{tabular}
\end{center}
\end{table*}

For each star we first find a solution for the non-magnetic wind by   running a first simulation is run with the magnetic field switched off. As the start solution defined by Eqs.~\ref{asymptotic} and \ref{inidens} is not in equilibrium near the star, the system undergoes some  adjustment before it settles in a stationary state. This takes about 200 ks for the \zetpup{} model. For other stars times vary dependent on the length scale set by the stellar radius and the terminal velocity. If the stationary state thus  found does not match the mass loss rate and terminal velocity required to meet observations we repeat the run with varying values for the $\alpha$ and $\bar{Q}$ parameters until we find a solution with the required properties. 
\subsubsection{O stars}
\begin{figure}
 \begin{center}
    \includegraphics[width=8.5cm]{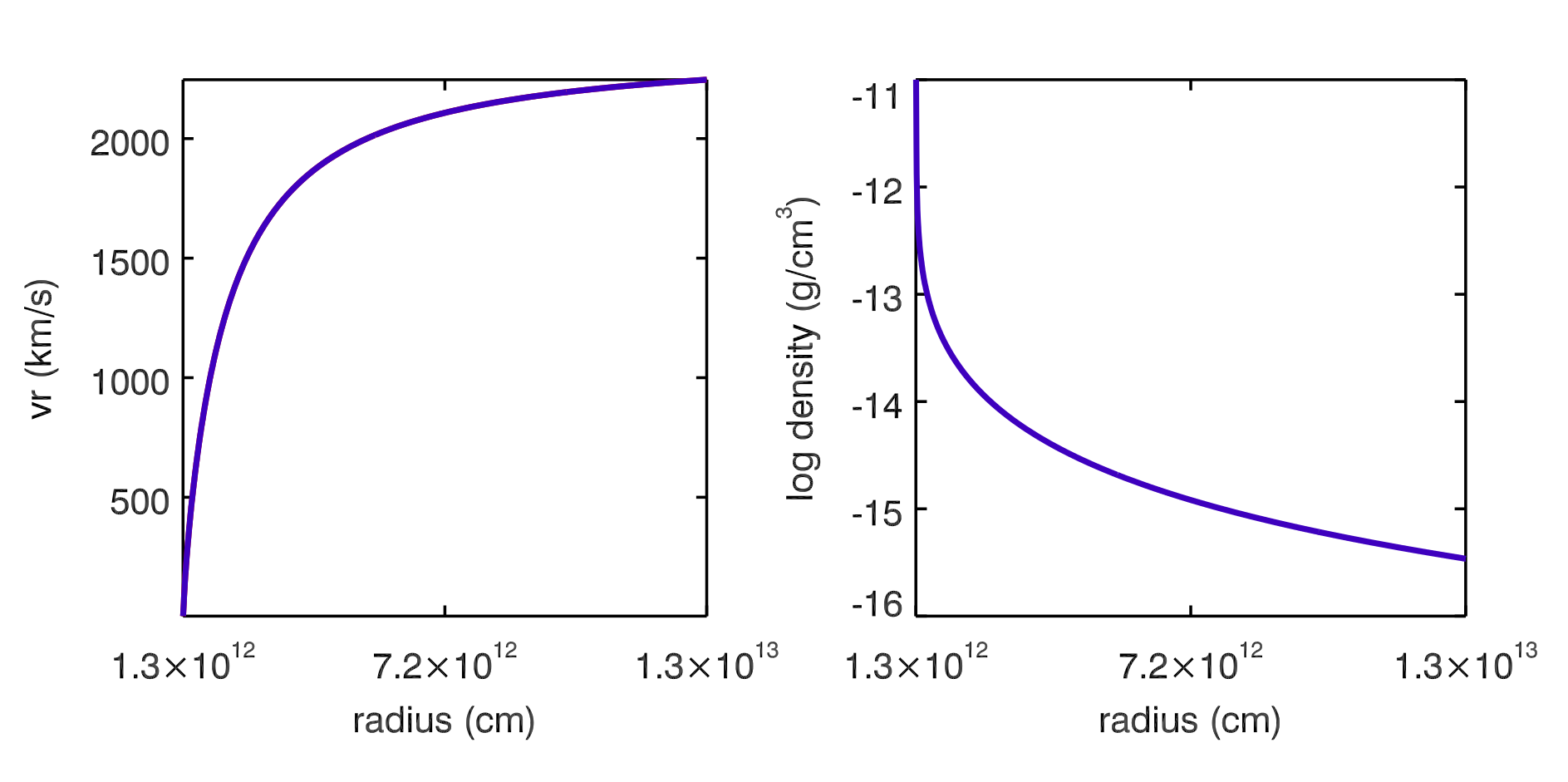}   
 \end{center}
 \caption{Radial velocity and density for $\zeta$ Pup with zero magnetic field ($\eta_*$=0). 
} 
 \label{eta0_wind}
\end{figure}

For a non-rotating and system without a magnetic field, spherical symmetry is preserved. 
Figure \ref{eta0_wind} shows the radial velocity and mass density as functions of radius for the evolved states for the \zetpup{} model. The density falls off by two orders of magnitude from its value of $4.3 \times 10^{-11}$ g/cm$^3$ on the inner boundary within a thin layer above the stellar surface and by another two orders of magnitude throughout the rest of the simulation box. The gas velocity increases correspondingly, reaching the sound speed of 20 km/s within the boundary layer at the inner boundary. At the outer boundary, the radial velocity is 2200 km/s. The mass loss rate is $2.6 \times 10^{-6}$ solar masses per year.

\begin{figure}
 \begin{center}
 { 
   \includegraphics[width=7.0cm]{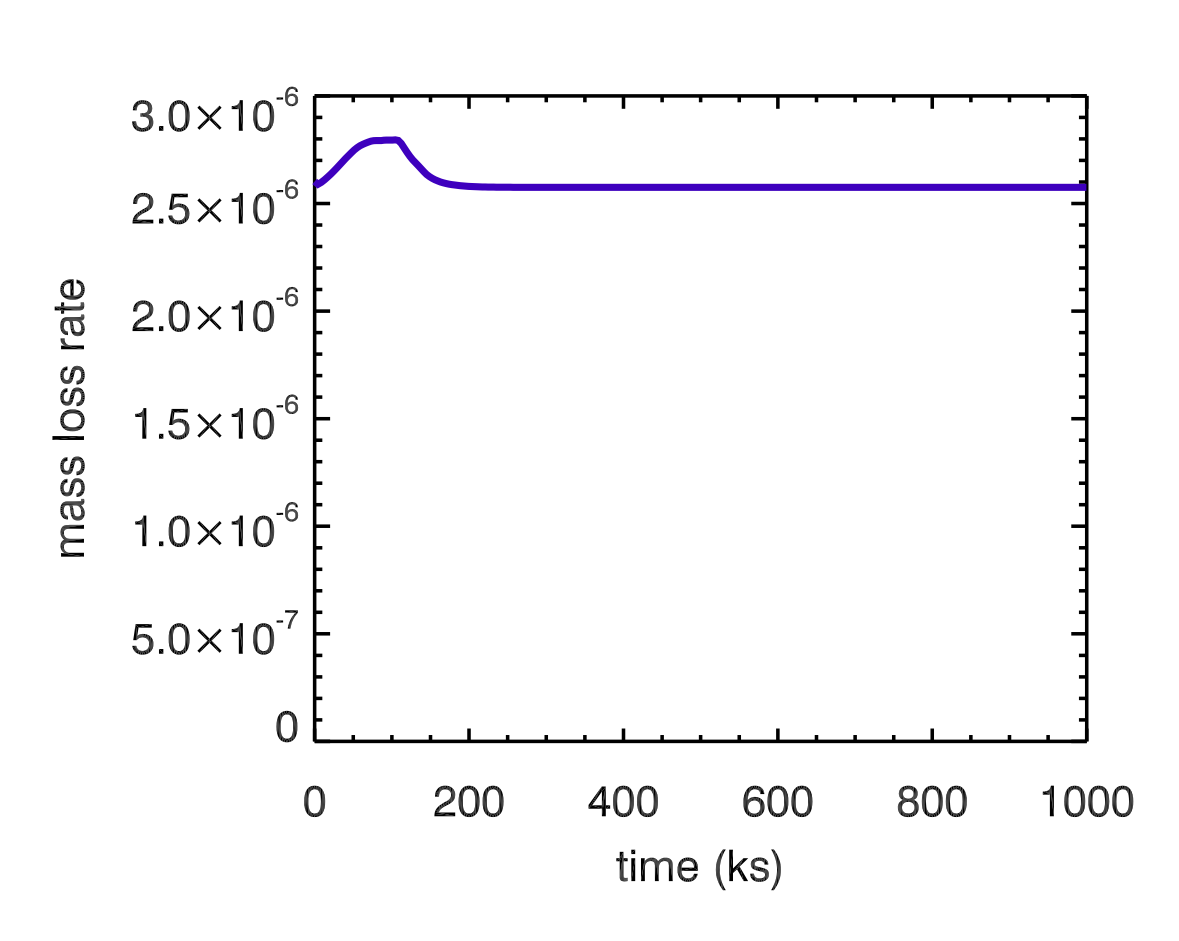} \\
   \includegraphics[width=7.0cm]{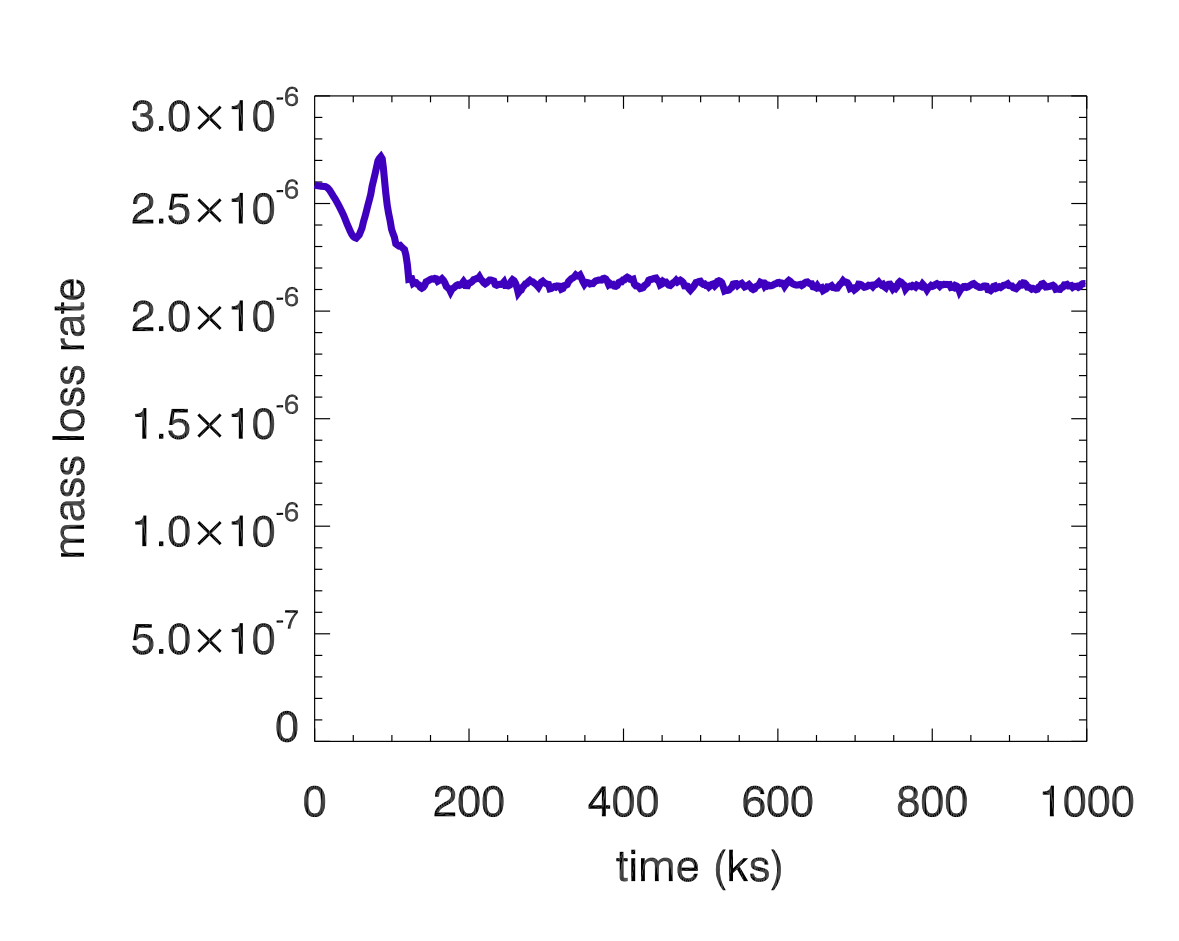} \\
   \includegraphics[width=7.0cm]{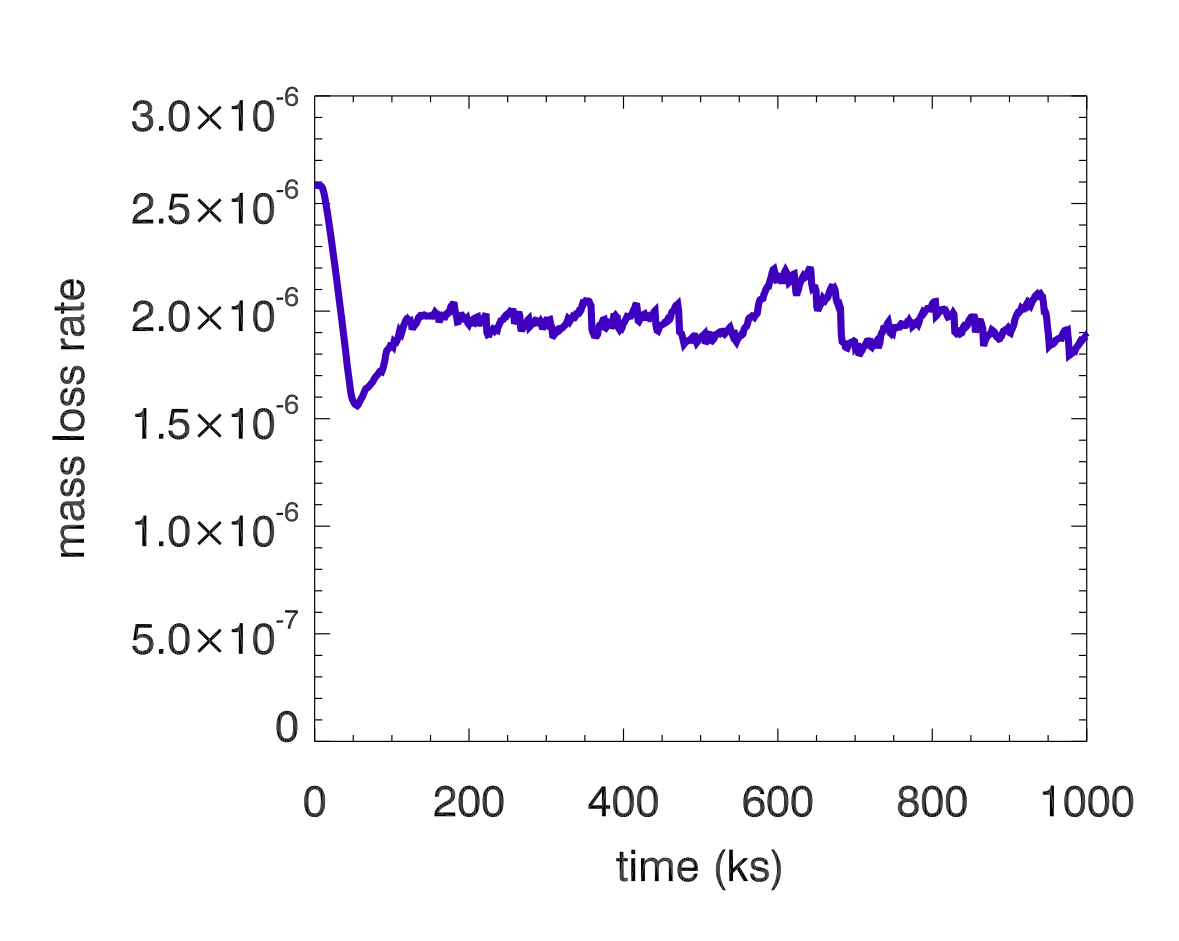}
   }
   \end{center}
 \caption{Time series of the mass loss rate for the \zetpup{} model with
          $\eta_*=0.1$ (top), 
           $\eta_*=1$ (middle),
           and $\eta_*=10$ (bottom). The unit on the vertical axis is solar masses per year.
           }
 \label{zetpup_momentum}
\end{figure}

Now the magnetic field is switched on and adjusted to meet the cases of weak ($\eta_*=0.1$), intermediate ($\eta_*$=1), and strong ($\eta_*$=10) magnetic field. Again, there is a transient phase of about 200 ks after which the system settles into a semi-stationary state. Figure \ref{zetpup_momentum} shows the mass loss rates as a function of time for the $\eta_*=$ 0.1, 1, and 10 cases. As the mass loss rate at any particular radius can show strong short-term variations, radial averages over the outer 3/4 (in radius) have been applied. 
In the weak field case, $\eta_*=0.1$, the gas flow actually becomes stationary. For stronger magnetic field the mass loss rate fluctuates. The time scale and amplitude of these fluctuations increase with field strength. In all three cases the mass loss is roughly steady after about 150 ks. The $\eta_*=1$ case shows only short time, small amplitude variations after that time. For stronger fields there larger amplitude variations on time scales up to 100 ks.

\begin{figure*}
  \begin{center}
  \mbox{
  \includegraphics[width=42mm]{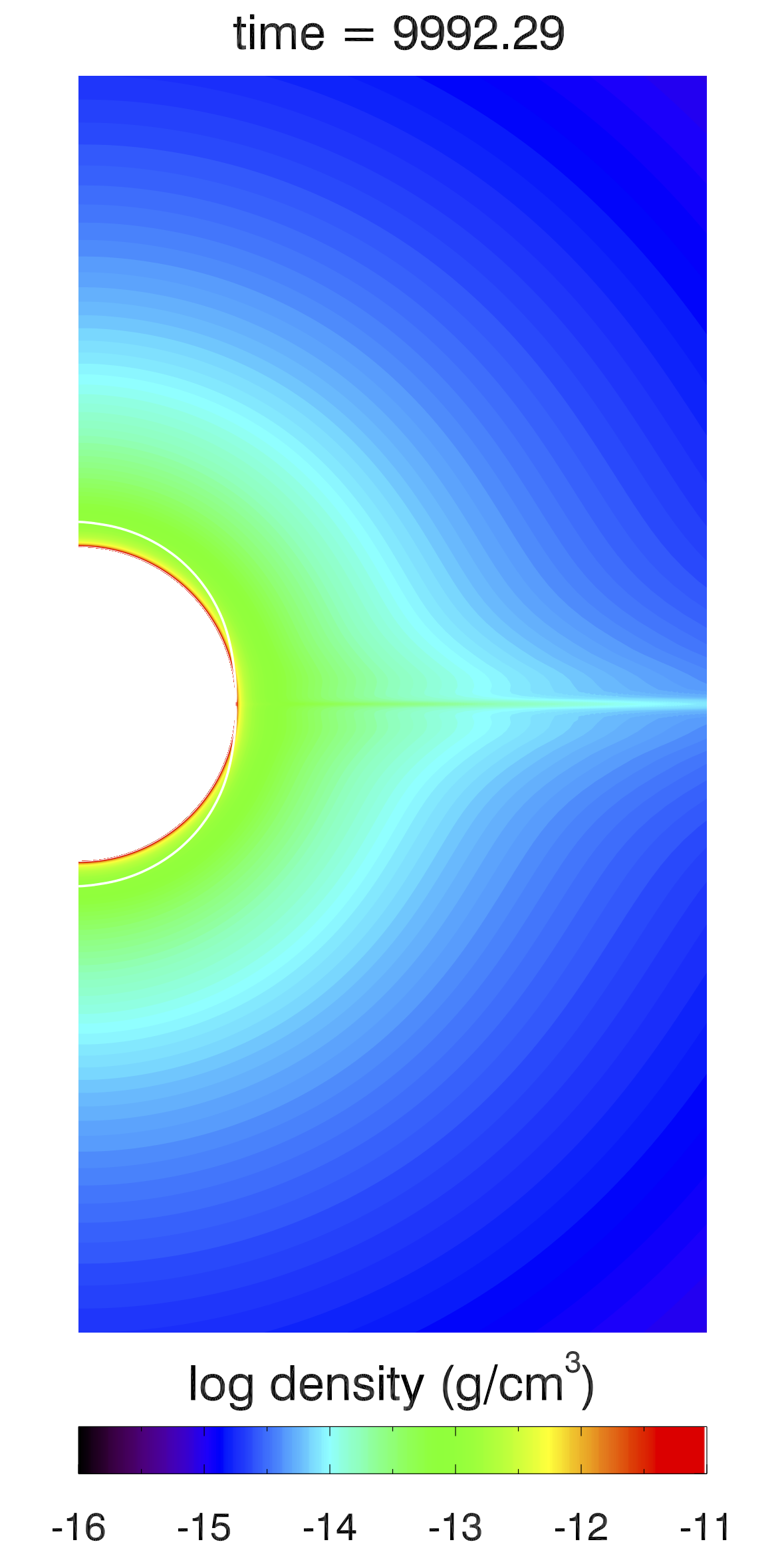}
   \includegraphics[width=42mm]{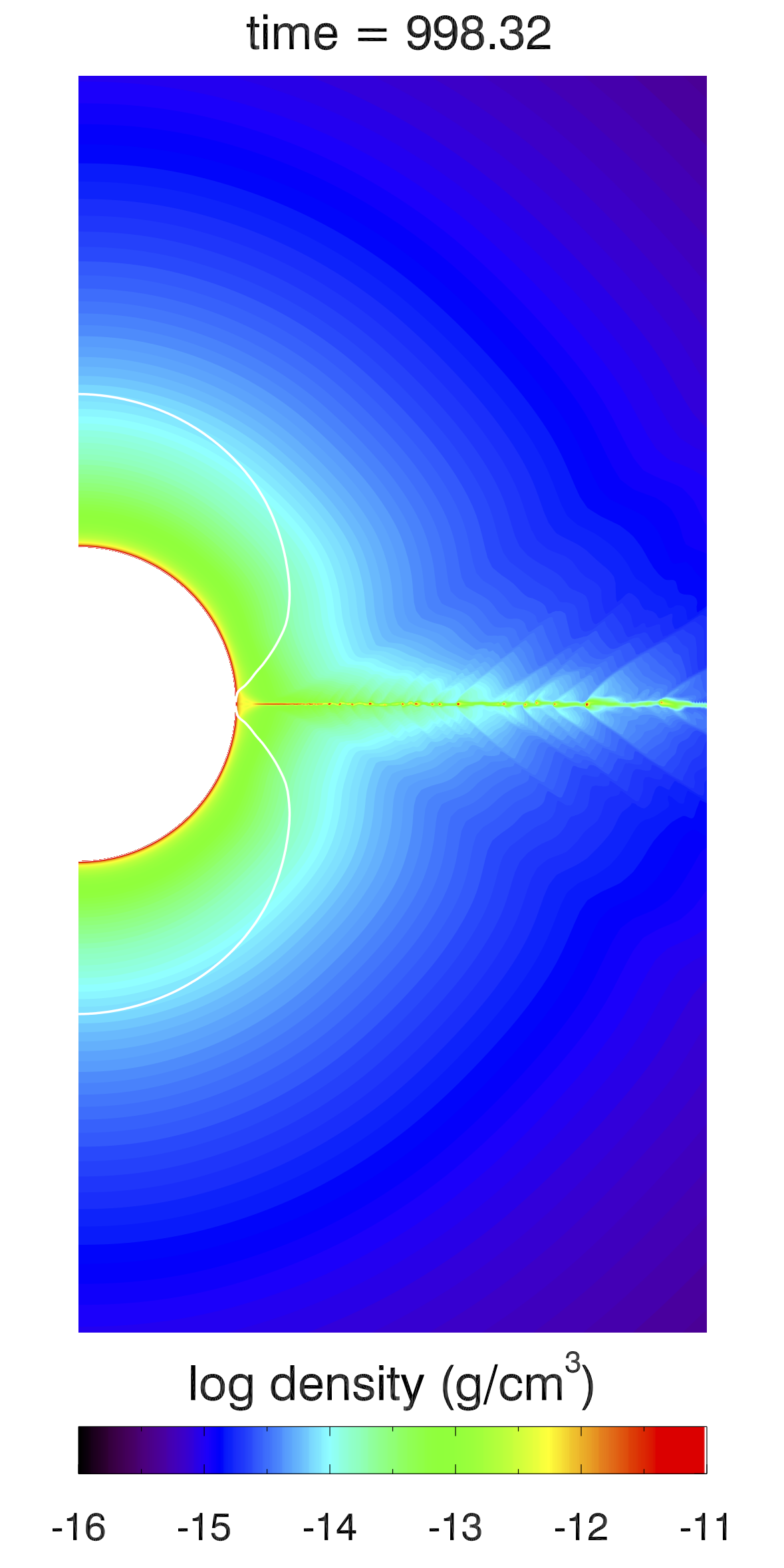}
    \includegraphics[width=42mm]{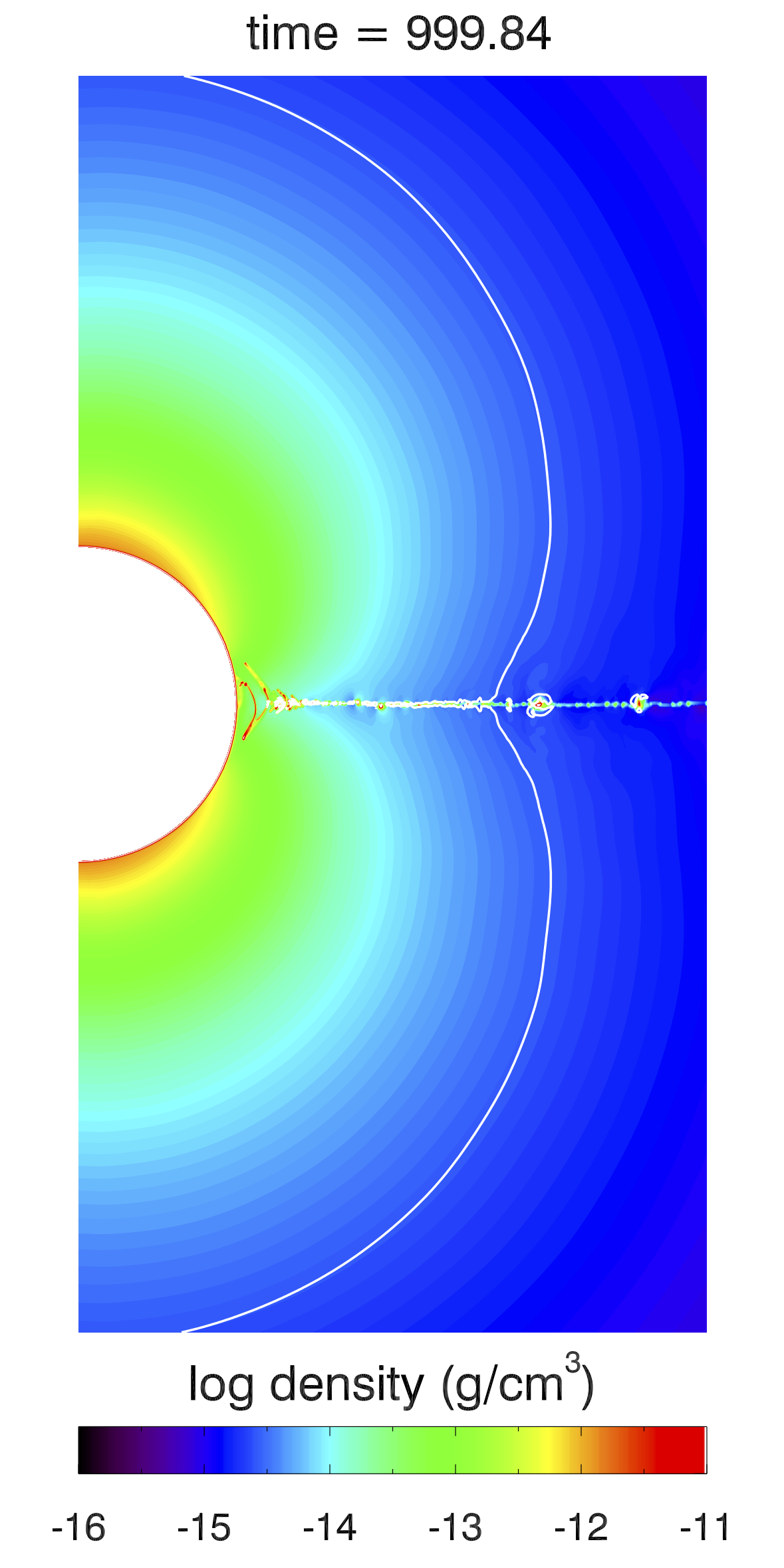}
    \includegraphics[width=42mm]{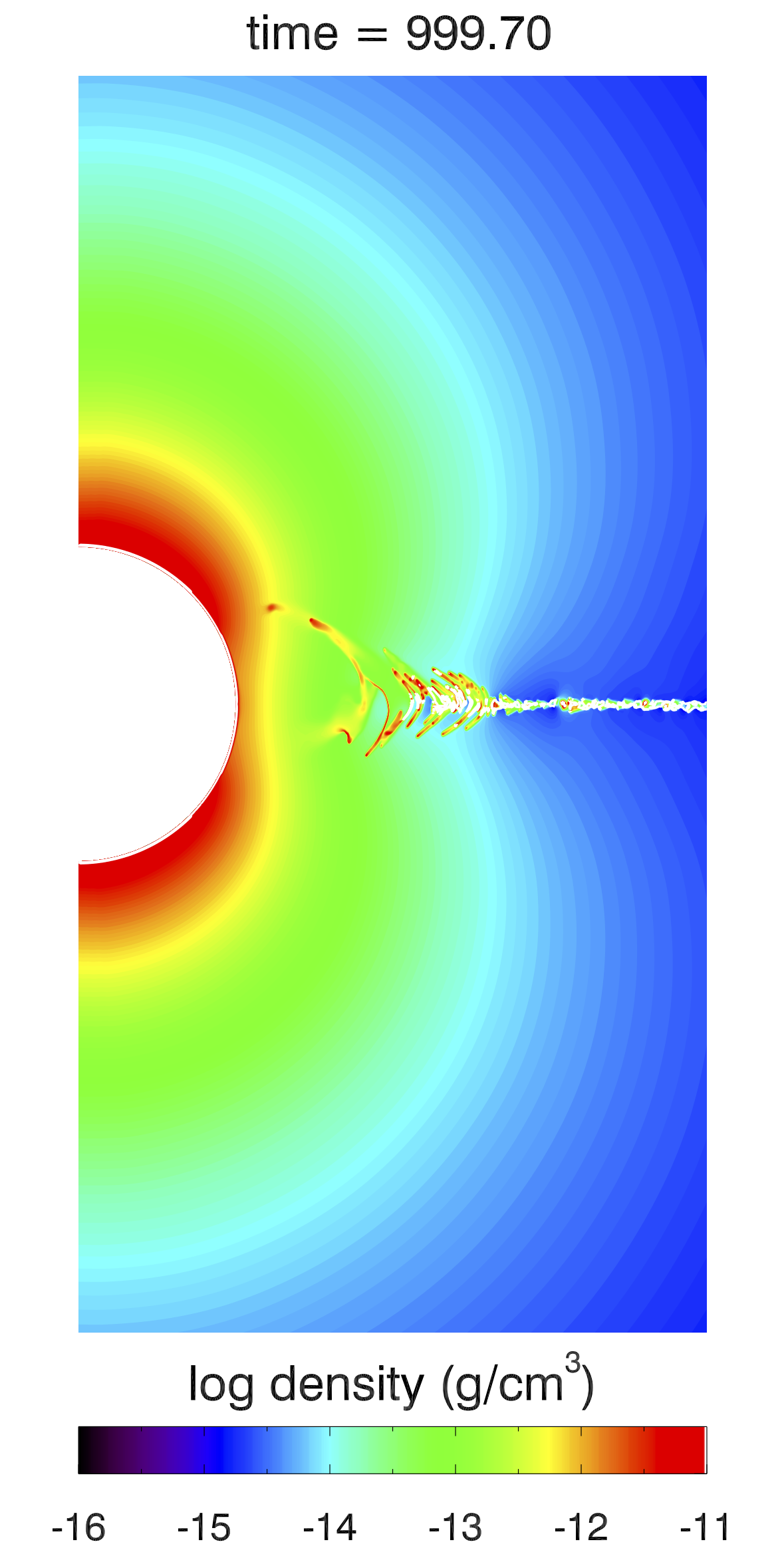}
    }
  \end{center}
\caption{Mass density in the vicinity of the \zetpup{} model for $\eta_*=0.1$, $\eta_*=1$, $\eta_*=10$  and $\eta_*=100$ (left to right). The unit of the time stamps is ks. The white line indicates $\eta=1$.
}
\label{zetpup_zoom}
\end{figure*}

Figure \ref{zetpup_zoom} shows snapshots of the density distributions. The boundary of the magnetically-dominated region is indicated by the black $\eta=1$ contour. 
\begin{figure*}
  \begin{center}
   \mbox{\includegraphics[width=110mm]{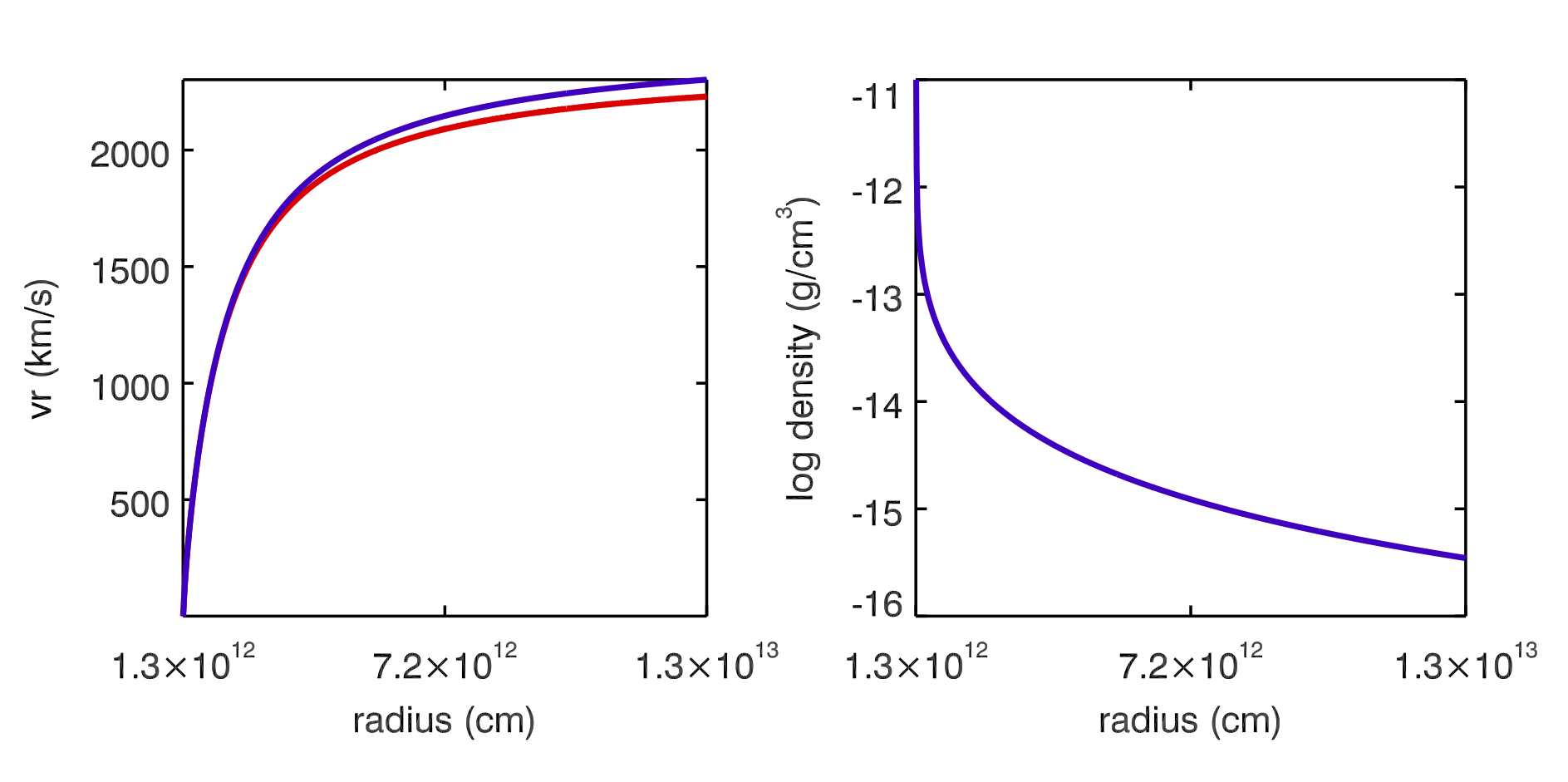}
   \includegraphics[width=55mm]{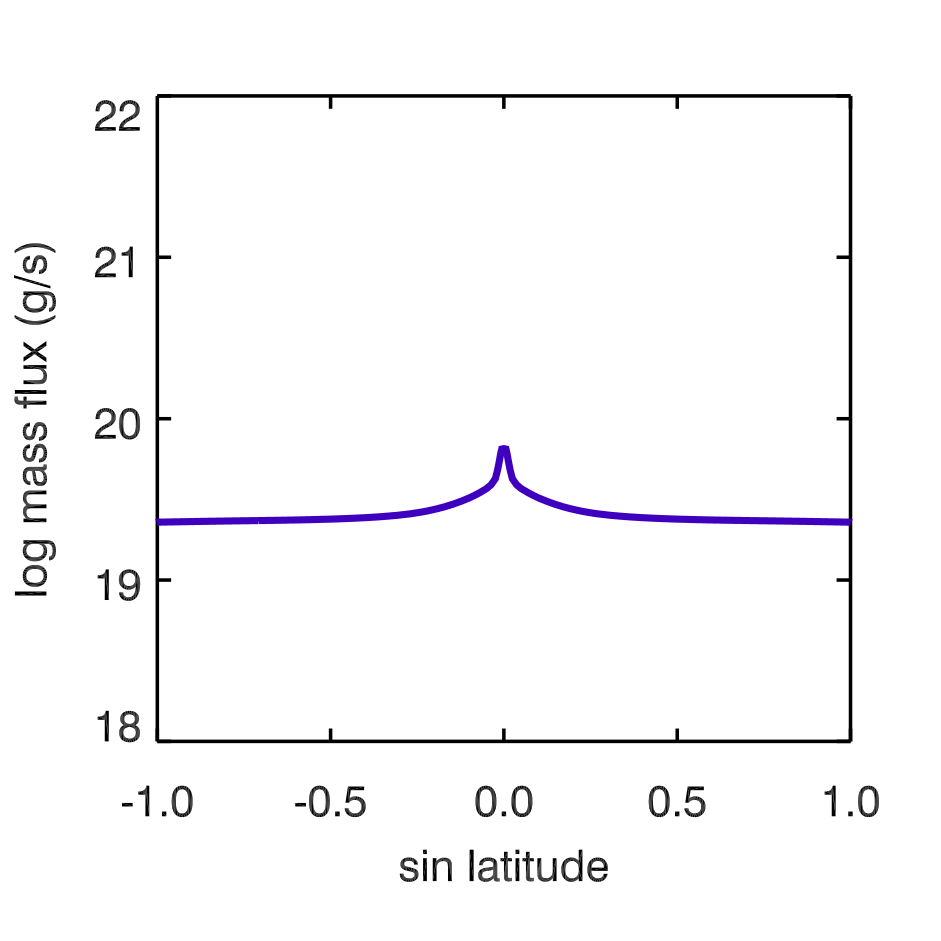} }\\
     \mbox{\includegraphics[width=110mm]{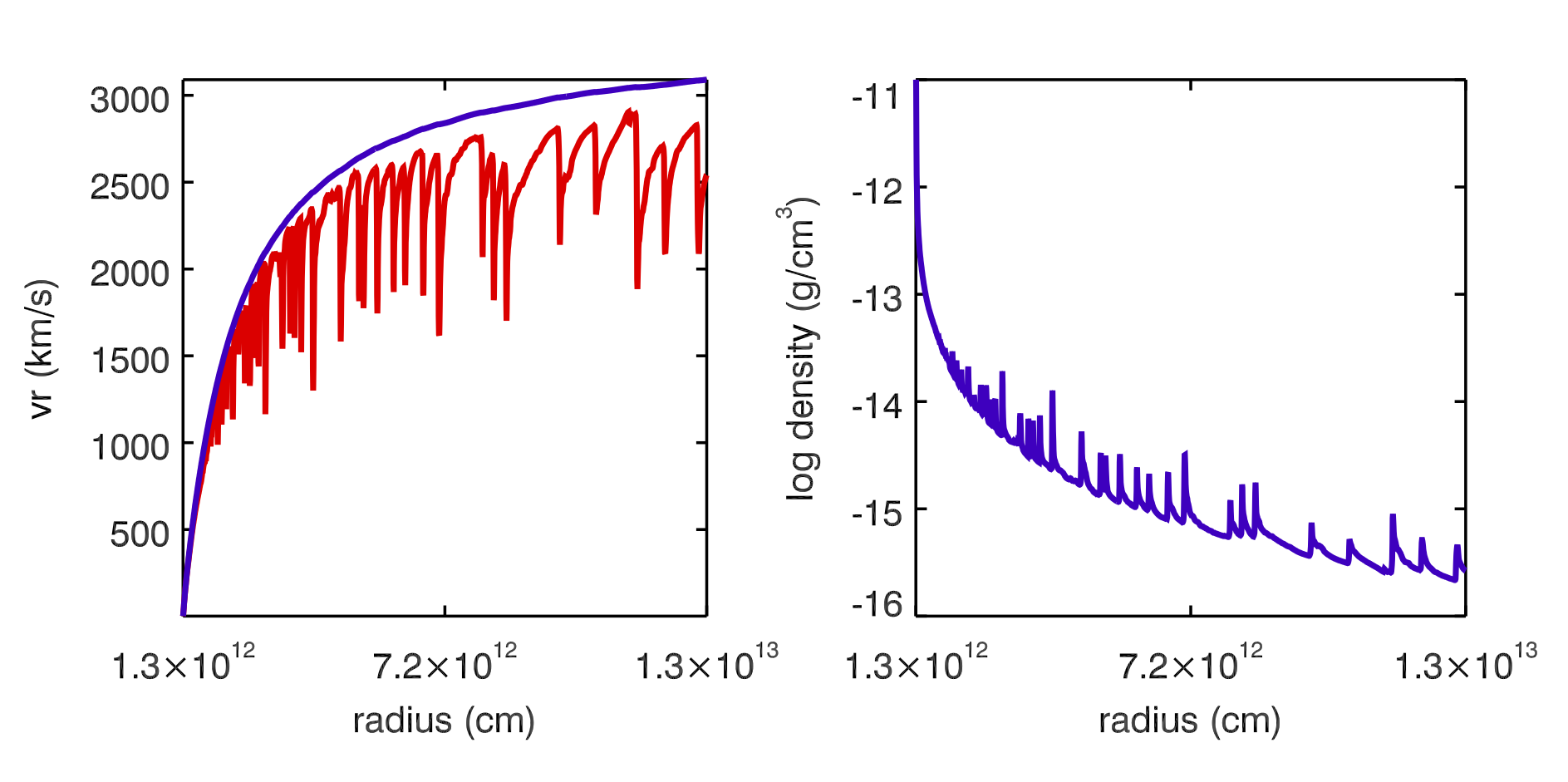}
   \includegraphics[width=55mm]{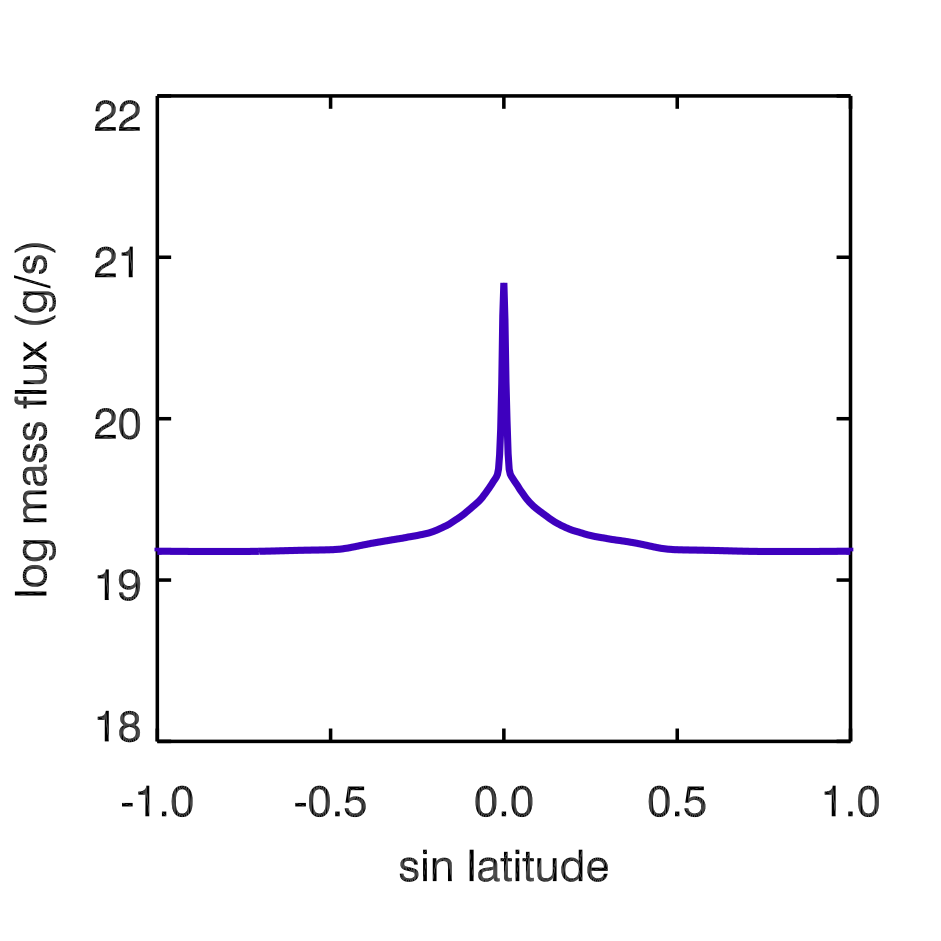}}\\
    \mbox{ \includegraphics[width=110mm]{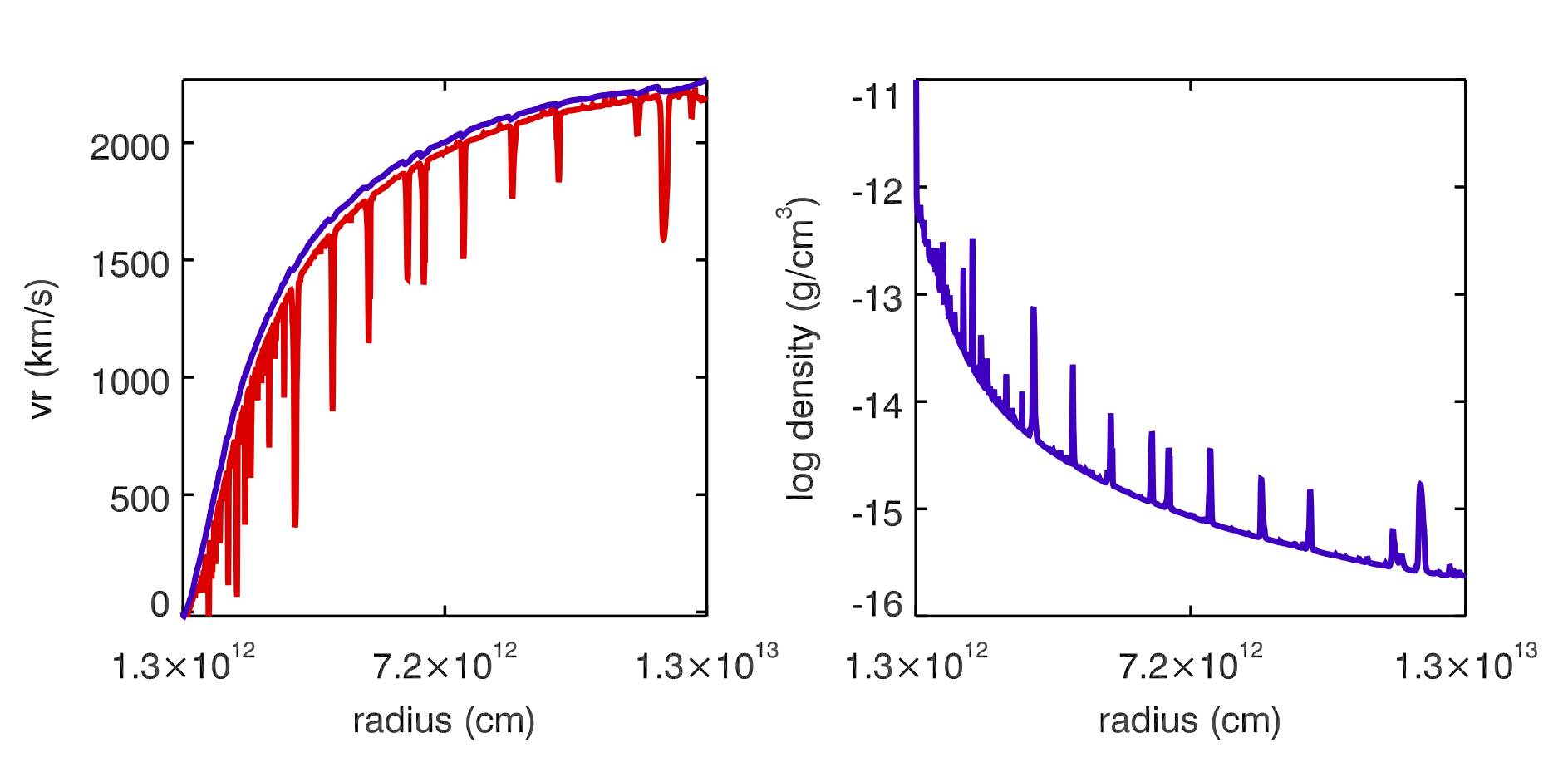}
   \includegraphics[width=55mm]{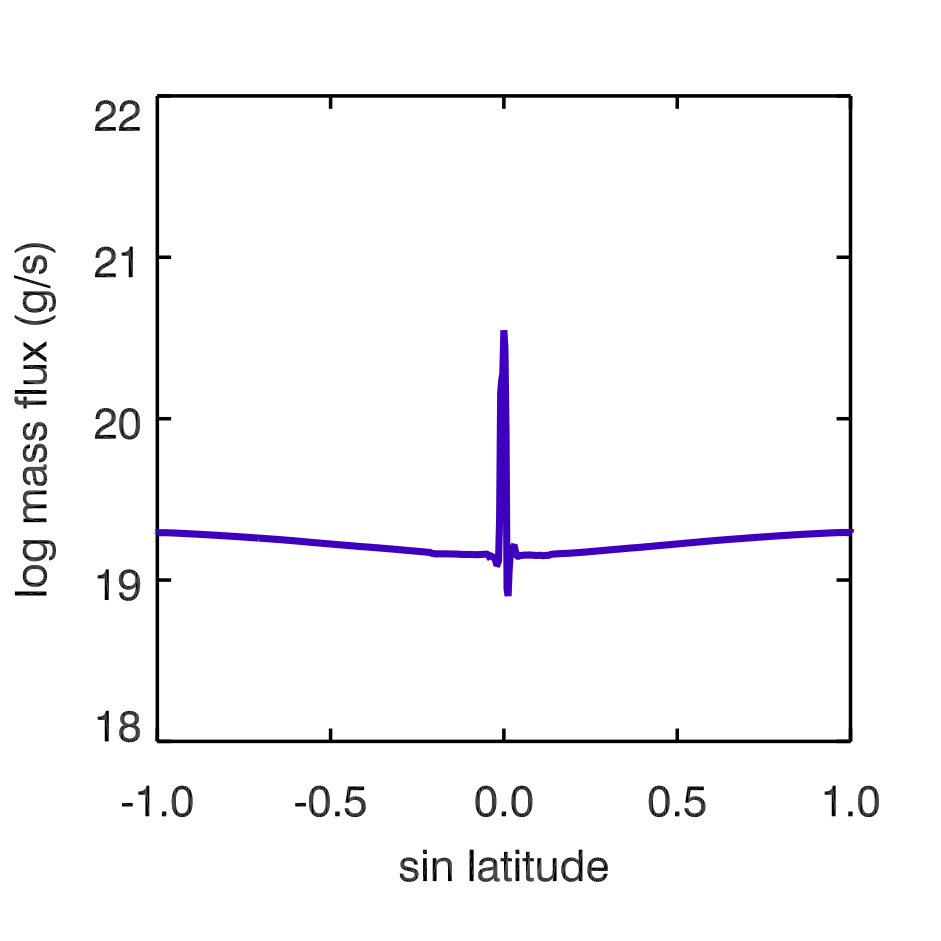} }
  \end{center}
 \caption{From left to right: latitudinally-averaged radial velocity vs.~radius, latitudinally-averaged gas density vs.~radius, and radially-averaged mass flux vs.~colatitude for $\zeta$ Pup. Top to bottom: $\eta_*$ = 0.1, 1,  10. Radius is in cm, velocity in km/s, and density in g/cm$^3$. The red lines show the latitudinally-averaged radial velocity with the density as a weight function.
 }
  \label{zetpup_winds}  
\end{figure*}
In the weak field case, that region is small but finite. With increasing field strength the size of the magnetically dominated region increases substantially. However, the magnetic field energy falls off more rapidly with increasing radius than than the kinetic energy of the wind. The outer part of the simulation is therefore always dominated by the wind. In the wind-dominated region, the initially dipolar magnetic field opens up and is stretched into a field configuration close to a split monopole, with a current sheet in the equatorial plane. In the strong field case, field lines remain closed at low latitudes near the stellar surface. Gas from the wind then gets trapped and falls back to the star. Magnetic field lines with foot points at high latitudes always open up at large distances.

In the weak field case, there is a noticeable flattening of the density isocontours throughout the simulation box. The gas flow is still smooth and essentially steady, though. 
In the equatorial plane a disk of enhanced mass density and reduced gas velocity forms.
This also shows in the mass flux shown in Fig.~\ref{zetpup_winds}, along with the radial velocity and density as functions of radius. As these functions are  not spherically symmetric for non-zero magnetic field, the plots show surface averages.
For the intermediate field strength, the gas disk in the equatorial plane is not smooth but consists of an intermittent series of gas rings separated by gaps. Density contours above the disk are slightly prolate now rather than oblate. 

The intermittency of the disk is even more pronounced in the strong field case, $\eta_*=10$. The $\eta=1$ contour has moved farther away from the stellar surface, except in and near the equatorial plane, where kinetic energy dominates in the gas rings and the magnetic field energy in the spaces in between. Near the stellar surface there is a region at low latitudes where the magnetic field maintains its original dipolar topology with closed field lines. Gas trapped in this region falls back towards the stellar surface along the field lines. 
The disruption of the inner disk by a region of closed field lines is much more pronounced for $\eta_*=100$ as shown in the right panel of Fig.~\ref{zetpup_zoom}.  Here the original dipolar field geometry is preserved in much larger region both in radius and latitude and the density distribution shows a filament structure that is aligned with the magnetic field.

The left and centre panels in Figure \ref{zetpup_winds} show the latitudinally-averaged values of the radial velocity and density, i.e.
 \begin{equation}
   \langle v_r \rangle =\frac{1}{2} \int_0^\pi v_r \sin \theta d\theta
 \end{equation}
 and likewise for $\rho$. The right panels show the radially-averaged mass flux
\begin{equation}
   F_m=\frac{1}{R_{\rm out}-R_{\rm in} } \int^{R_{\rm out}}_{R_{\rm in}} \rho v_r r^2 dr
\end{equation}  
as function of latitude.  
 While the latitudinally-averaged velocity shows a smooth increase with radius, the density shows a sawtooth-pattern for intermediate and strong confinement. This is the contribution from the disk and reflects the fragmented structure of the latter. Each spike in the density represents a ring of increases mass density. The latitudinally-averaged radial velocity does not show this behaviour because the disk volume is a very small fraction of the total volume. 
The red lines show the latitudinally-averaged radial velocity where the averaging is done with the density as a weight function,
 \begin{equation}
   \bar{ v_r } =\frac{1}{\langle \rho \rangle} { \int_0^\pi v_r \, \rho \sin \theta d\theta}.
 \end{equation}
 In the weak field case the density-weighted velocity is just a little reduced because of the contribution from the disk. The intermediate and strong field cases
 show a sawtooth pattern similar to $\langle \rho \rangle$ but pointing downwards.
 Note that for strong confinement the radial profiles of radial velocity and mass density are substantially changed from the non-magnetic case. The increase of the radial velocity is much slower and the mass density much larger close to the star. The increase of the mass density with increasing magnetic field strength is also evident in Fig.~\ref{zetpup_zoom}, which shows an increase over the weak field cases particularly at high latitudes.      
  
The right panel  in Figure \ref{zetpup_winds} shows the deviation of the mass flow from spherical symmetry. In the weak field case, there is just a small increase at low latitudes while the intermediate and strong field cases show a sharp peak in the equatorial plane. 
Mass loss is thus enhanced in the disk, a the increased density outweighs the reduced speed. The total mass loss is never dominated by the disk, though. Outside the disk the mass flux shows a slight increase with latitude for strong fields while the weak field case shows the opposite trend.
 While the radial velocity at the outer boundary first increases from 2200 km/s to about 3000 km/s and then decreases to about 2200 km/s as the field strength increases, we do not find a dependence of the mass loss rate on the field strength. The radially-averaged mass loss rate (computed over the outer half (in radius) of the simulation box shows values of about $(2.6 \pm 0.1) \times 10^{-6} M_\odot/a$ and the variation between snapshots from different simulation runs is not larger than between snapshots from the same run. 

\tetori{} is a massive star of spectral type type O7Vp \citep{sota11} for which \cite{donati02} detected a surface magnetic field of dipolar geometry with a polar strength of about 1.1kG. The effective temperature and radius listed in Table \ref{parameters} have been adopted from \cite{simondiaz06} the mass-loss rate from \cite{howarth89}, the terminal speed from \cite{stahl96}.
The star has recently been discovered to be a binary. \cite{balega15} find a mass of 33.5 \msol{} for the primary component. \cite{babel97b} applied their MCWS model to this star to explain the observed X-ray emission and concluded that a dipole magnetic field with a strength of at least 270-370 G is necessary for their model to apply. \cite{gagne05} used numerical simulations with the setup from \cite{udd02} to model the X-ray emission of $\theta^1$ Ori C with a value of 7.5 for the confinement parameter. With the parameters listed in Table \ref{parameters}, we arrive at a value of 24.4. 

\begin{figure*}
 \begin{center}
   \mbox{
    \includegraphics[width=42mm]{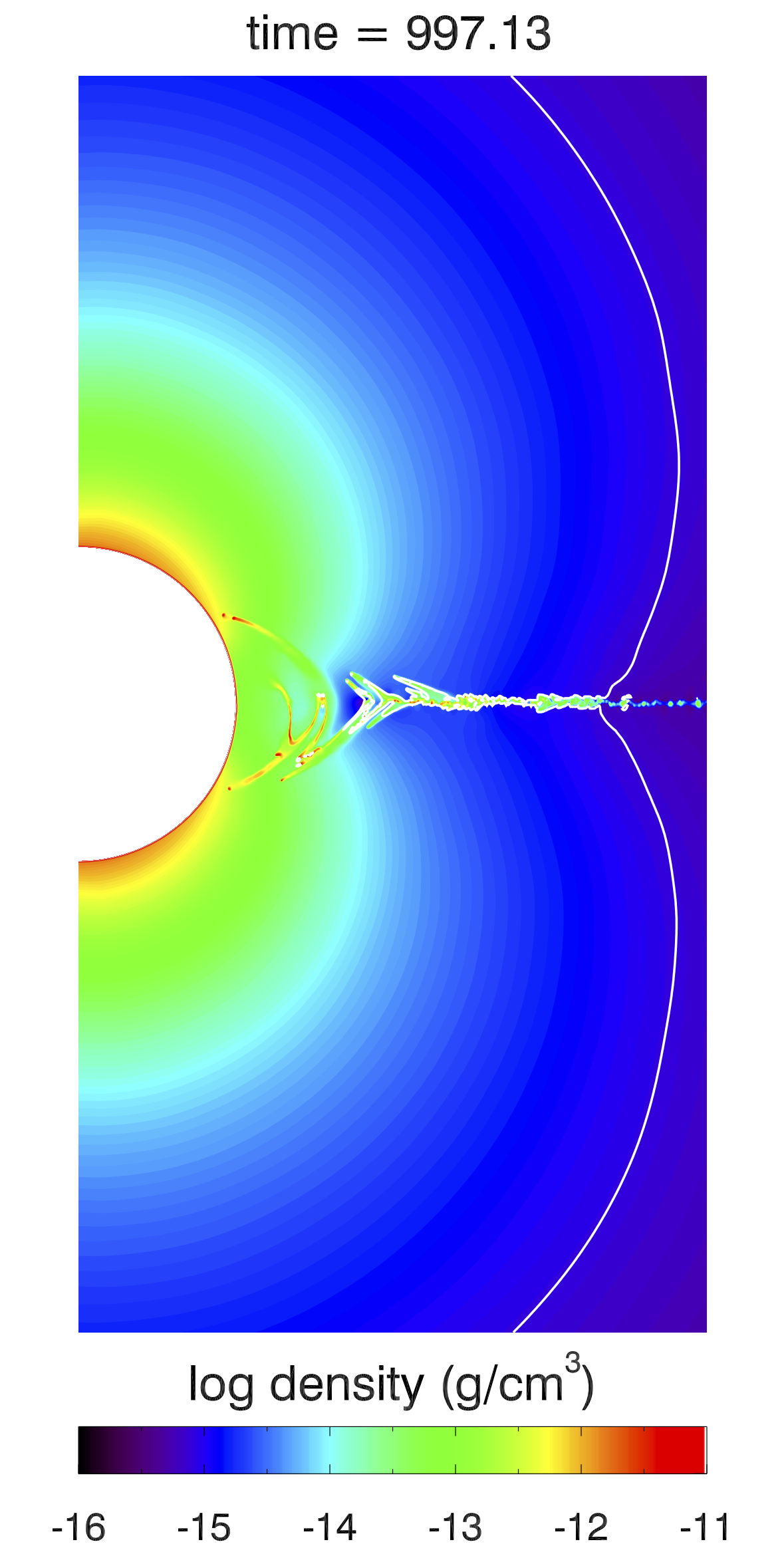}
    \includegraphics[width=42mm]{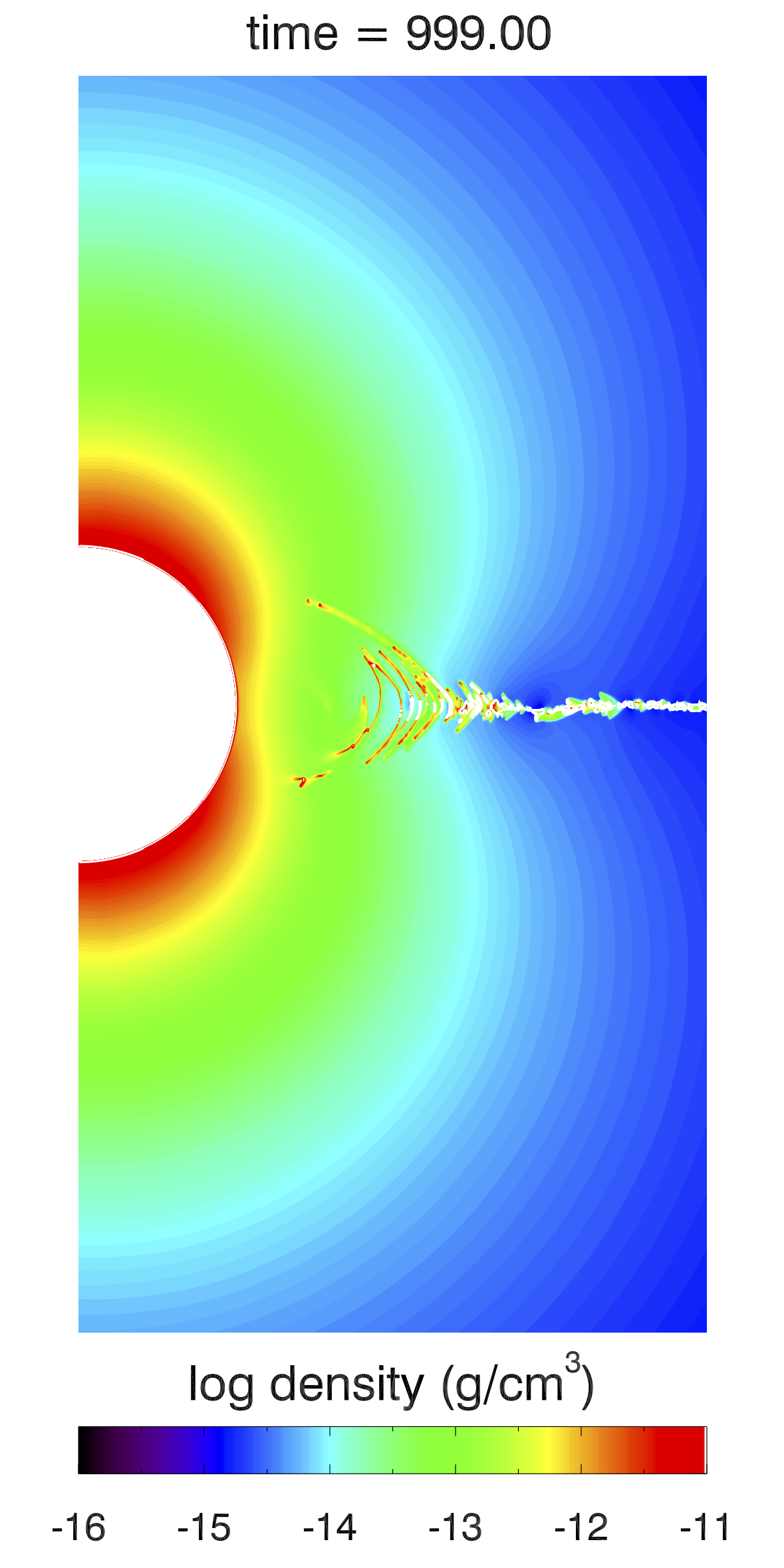}
      \includegraphics[width=42mm]{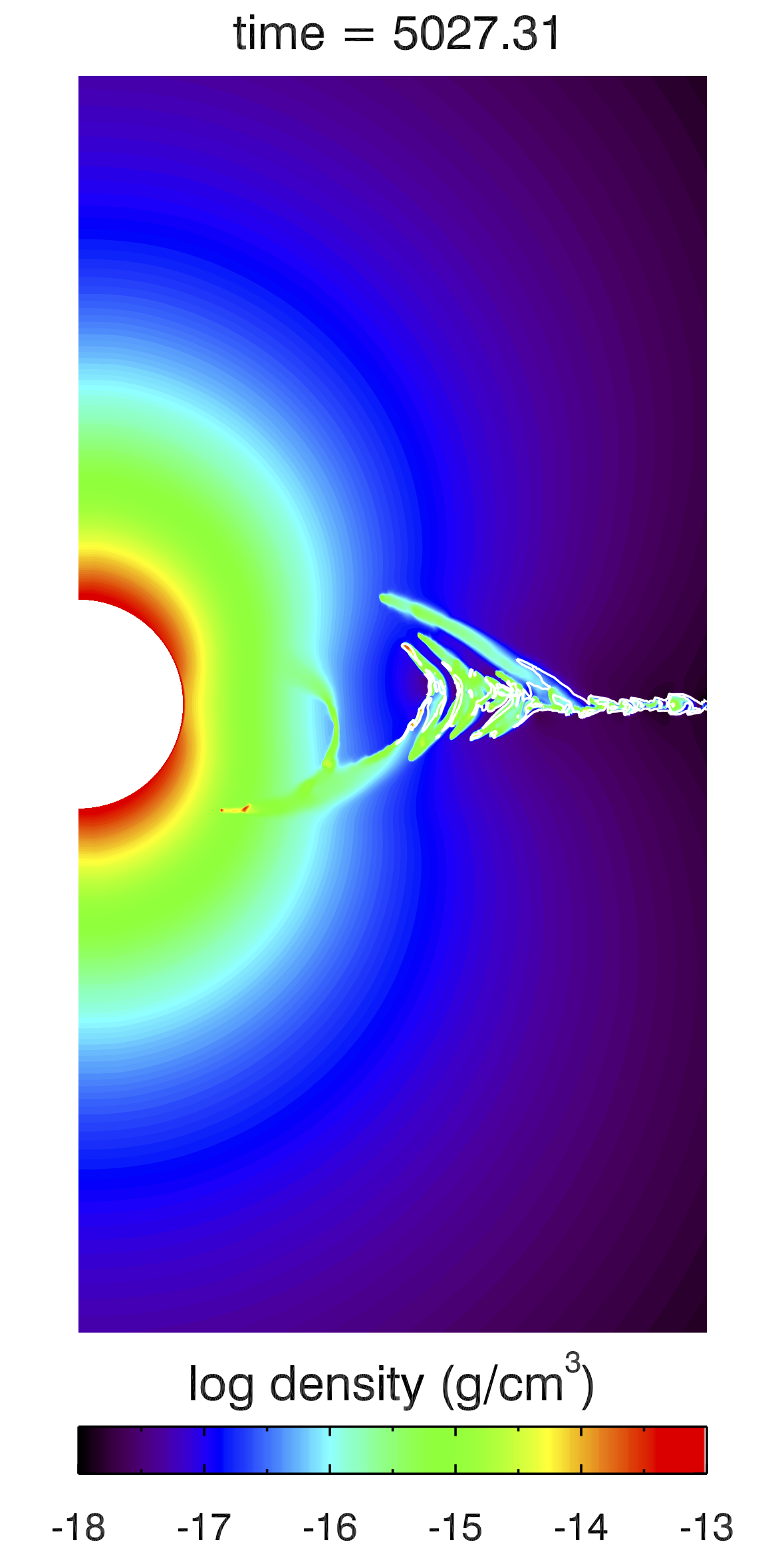}
     \includegraphics[width=42mm]{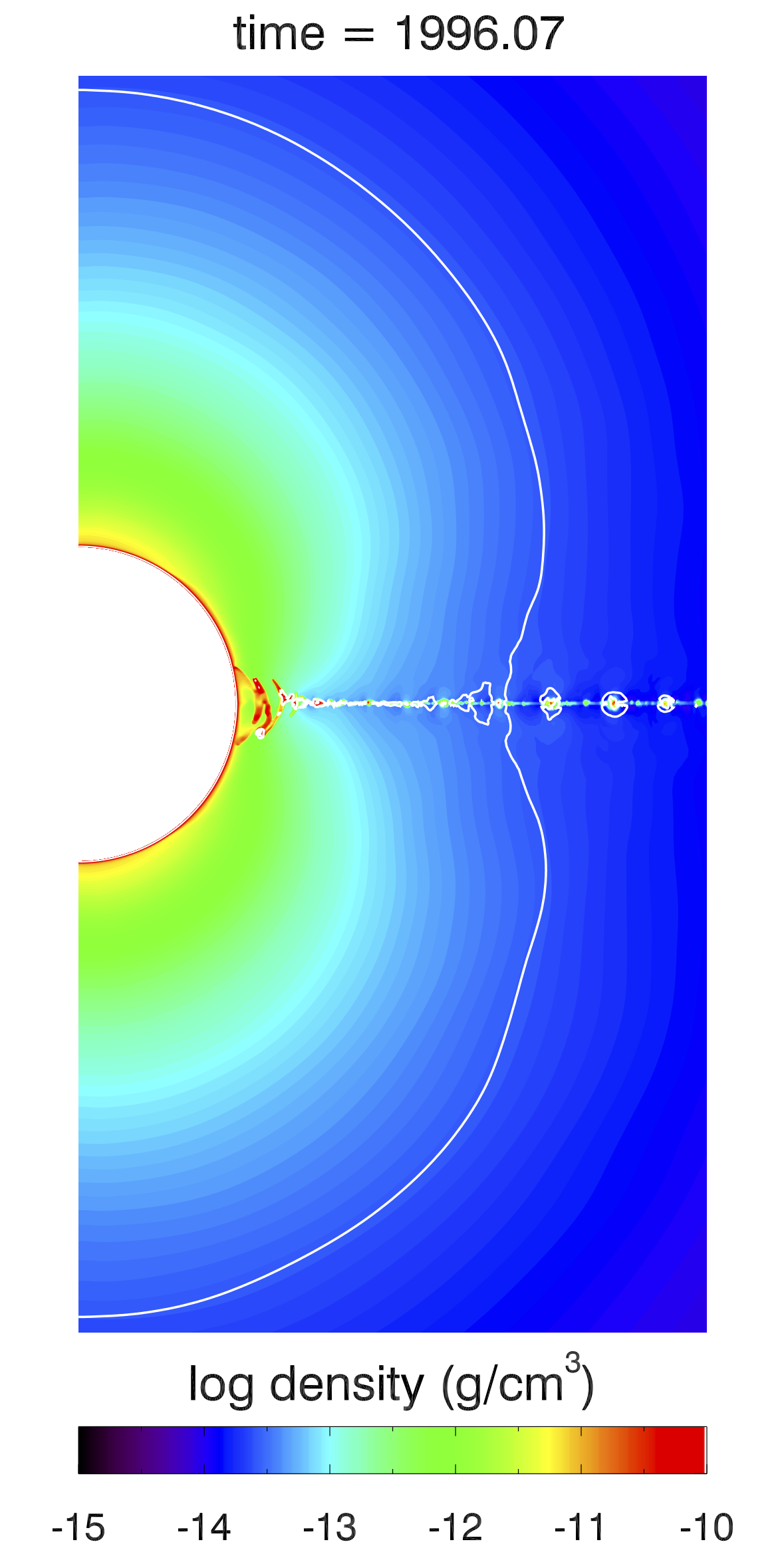}   
    }
 \end{center}
 \caption{Left: Mass density distribution around the star for  \tetori{}, \cpddd{}, \sigori{}, and WR105.}
 \label{tet1}
 \end{figure*}

The left panel in Figure \ref{tet1} shows the density distribution around the star 
for $\eta_*$= 24.4. As expected for this value of \confin{}, it generally resembles the strong confinement cases of \zetpup{} model, with size of the closed field line region being substantially larger than the \confin{}=10 case but smaller than the \confin{}=100 case. 

\cpddd{} is an Of?p star for which a surface magnetic field has been detected.  
 \cite{hubrig13} infer  field strengths of $381\pm122$ G from all lines and $534\pm167$ G from hydrogen only. \cite{wade15} find a cyclic variation of the maximum field strength between 565 G and -335 G but argue that the stellar magnetic field is a tilted dipole with a maximum field strength of 2.6 kG. With the stellar and wind parameters as listed in Tab.~\ref{parameters} the latter field strength correspond to a values of 90 for the confinement parameter $\eta_*$ while the value of 534 G only yields a value of 4, corresponding to rather weak confinement.
The second panel (from left) in Figure \ref{tet1} shows our model for \cpddd{} for $\eta_*=90$. With the exception of the equatorial disk, the whole region shown is magnetically dominated, i.e.~lies inside the $\eta=1$ contour. As expected for the larger value of the confinement parameter, the closed field line region is more extended than for \tetori{}. Mass density in the immediate vicinity of the star is also higher. 
\subsubsection{\sigori{}}
\begin{figure}
 \begin{center}
    \includegraphics[width=70mm]{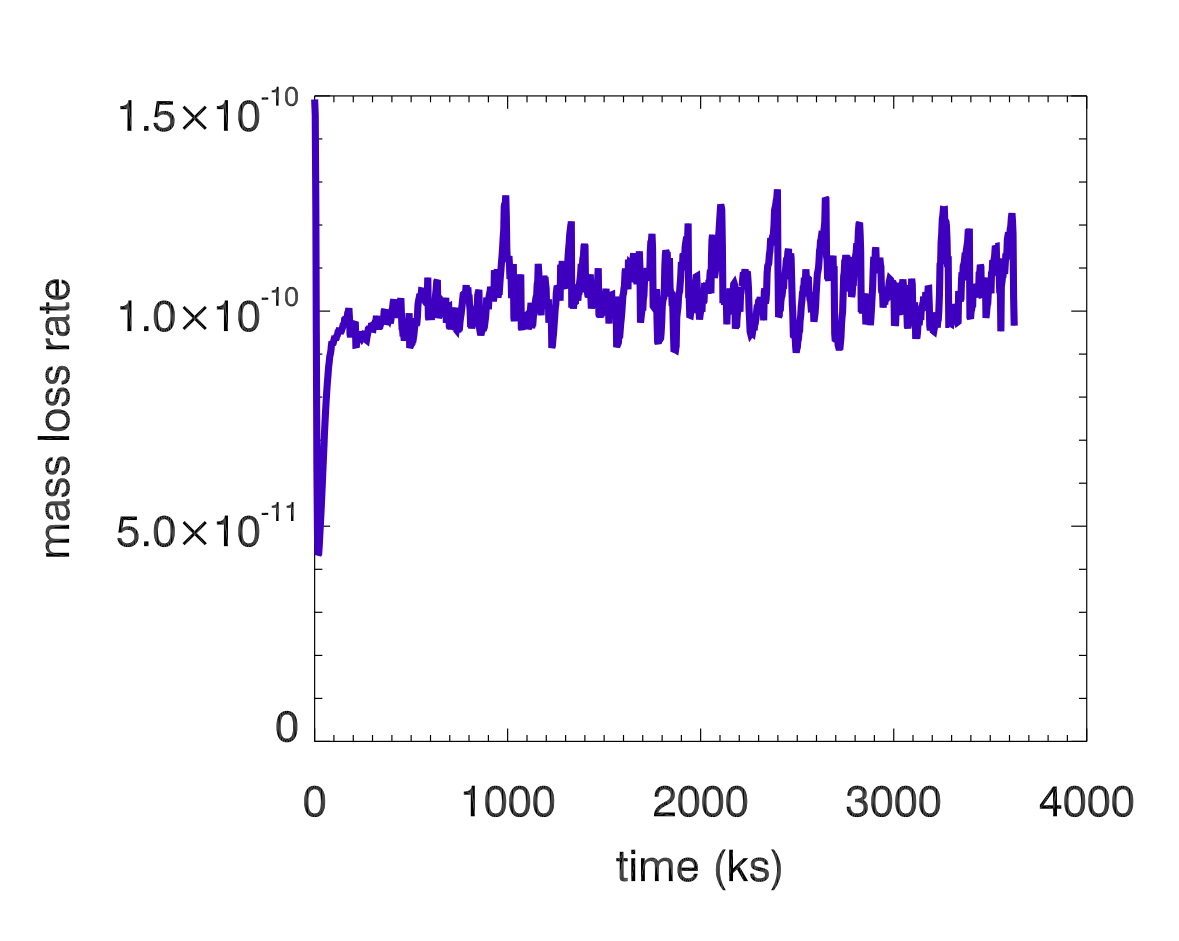}
    \includegraphics[width=70mm]{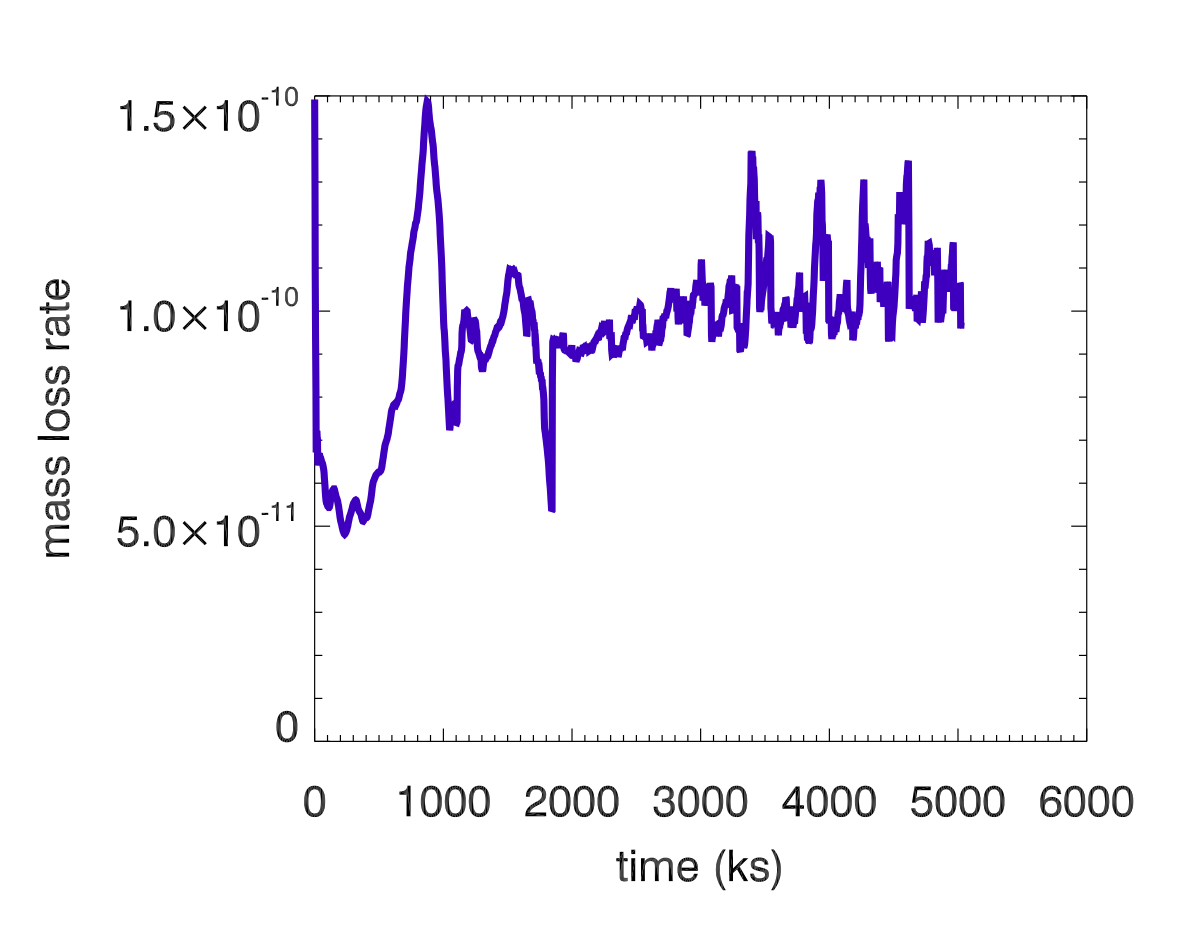}
 \end{center}
 \caption{Mass loss rate in solar masses per year vs.\,time for the \sigori{} model with $\eta_*=10^3$ (top) and $\eta_*=10^4$ (bottom.)}
 \label{sigmass}
\end{figure}

As an example of very strong confinement, we include the B2Vp star \sigori{}. On this star, a large-scale magnetic field with a polar strength of
about 10 kG has been observed \citep{landstreet78, groote82}. As this star combines a very strong surface magnetic field with a mass loss rate that is several orders
of magnitude lower than for the O stars (cf.\,table), a value of about $10^6$ results for the confinement parameter. 
With a rotation period of 1.19 days, the star is a fast rotator. The dipole axis is, however strongly tilted against the rotation axis and a treatment of the stellar rotation required a 3D setup. \cite{townsend05a} used the rigidly rotating magnetosphere (RRM) model by   \cite{townsend05b} to model the
magnetosphere of \sigori{}. This model, however, assumes a fixed magnetic field geometry and no back reaction from the gas. 
\cite{udd08} studied the effects of rotation on the magnetospheres of massive stars for the case of an aligned dipole. 
We defer the effects of rapid rotation to a future study and focus on the effect of strong magnetic confinement. 
The third panel (from left) in Figure \ref{tet1} shows the resulting density distribution for  $\eta_*=10^4$. The latter, which corresponds to a magnetic
field strength of 950 G, already shows an extended magnetosphere, with the Alfv\'en radius  lying outside the normal box radius of ten stellar radii. The simulations shown here use an outer radius of 20 stellar radii. Increasing the confinement parameter farther towards the real value of $10^6$ would 
have led to prohibitively long run times through smaller time steps and the need for even larger simulation boxes. However, Figure \ref{tet1} shows
that our setup is capable of modelling cases of very strong confinement. In both cases shown, large quantities of gas are trapped by the magnetic field
and fall back to the star. The mass loss rates are therefore substantially reduced. Figure \ref{sigmass} shows the mass loss rate vs.\,time. In both cases
the mass loss rate has been integrated over latitude and averaged over radius. As there is a substantial inward flow near the star, the averaging has been carried out over the outer half of the simulation box. Both cases show a strong decrease early in the simulation run and a recovery and saturation
at about $10^{-10}$ solar masses per year. The $\eta_*=10^4$ case shows stronger variation and takes significantly longer to recover from the initial drop.
%
\subsubsection{Wolf-Rayet stars}
We now run similar models for our sample of Wolf-Rayet stars. The $\bar{Q}$ and $\alpha$ parameters are chosen to match the observed mass loss rates and terminal velocities. For each star magnetic field strengths are chosen to match values of 0.1, 1, and 10, respectively, for the confinement parameter $\eta_*.$ The simulations are run for time spans from 200 to 2000 ks, chosen for each star to cover a time span about an order of magnitude longer  than it takes the wind to cross the simulation box and thus long enough for the system to settle into a quasi-stationary state.

The right panel in Figure \ref{tet1} show a snapshot of the density distribution near the end of the run for the star WR105 with $\eta_*=10$.
The outer boundary is at ten stellar radii. The plots show the inner region from x=0 to x=5$R_*$ and y=-5$R_*$ to y=5$R_*$. 

The $\eta=0.1$ cases look qualitatively very similar to the corresponding case for the $\zeta$ Pup model, showing the same flattening.
For $\eta_*=1$ both models show more pronounced flattening of the density contours and a disk in the equatorial plane. The disk is distinctly fractured and  blobs of enhanced density in the disk cause wake pattern in the wind above the disk. 
For $\eta_*=10$ the disk is even more fractured and there is a gas flow along the magnetic field lines at low latitudes close to the stellar surface. Density contours are now distinctly prolate rather than oblate, with a pronounced dent near the equatorial plane. The wake pattern observed for $\eta_*=1$ has largely vanished.
%
%
%
\begin{figure*}
 \begin{center}
 \mbox{
 \includegraphics[width=50mm]{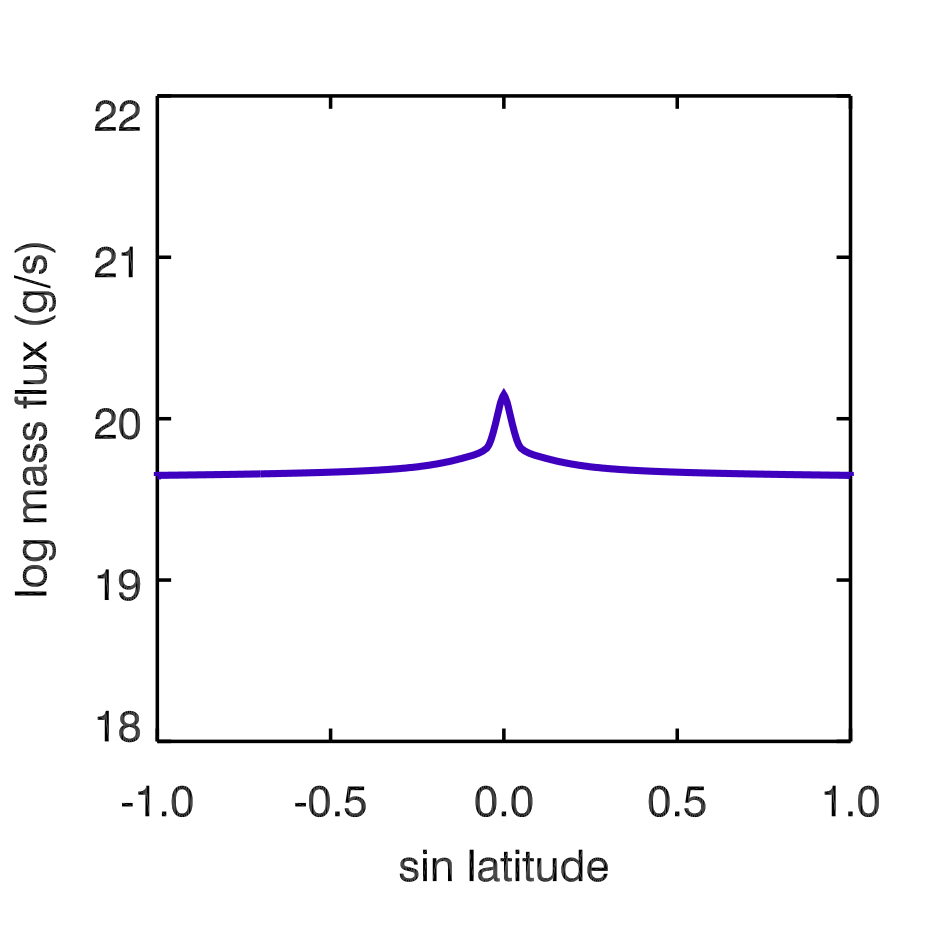}
 \includegraphics[width=50mm]{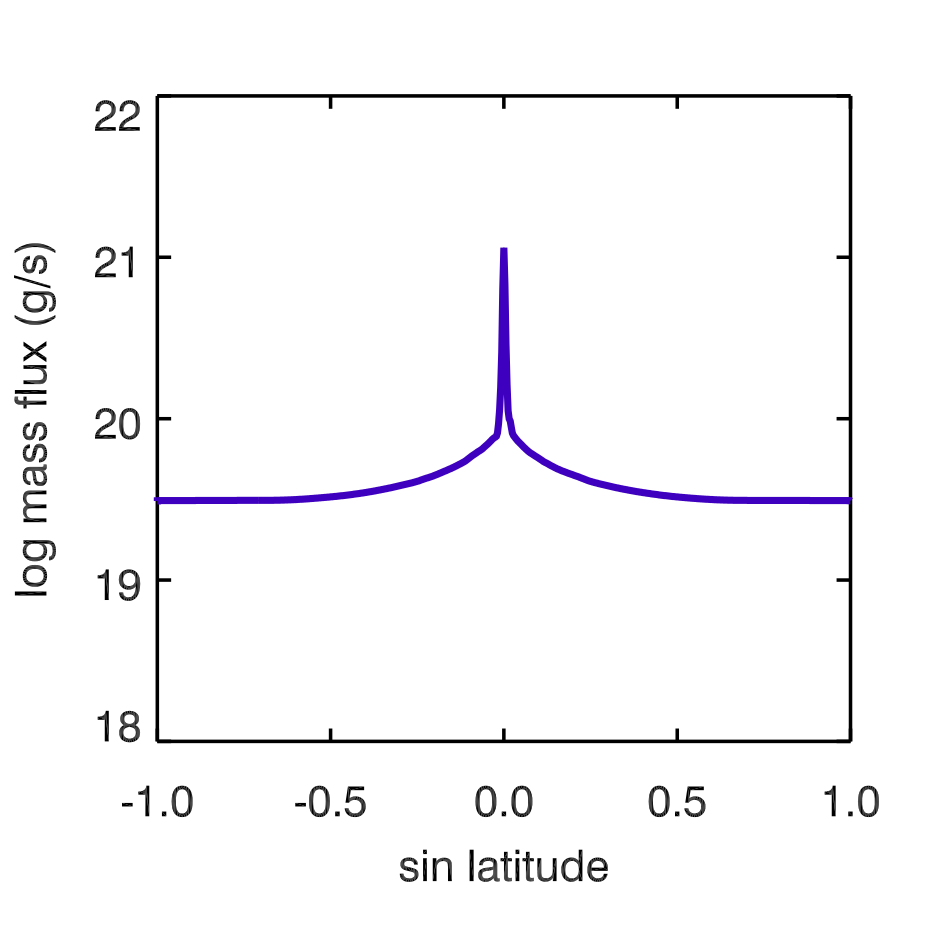}
 \includegraphics[width=50mm]{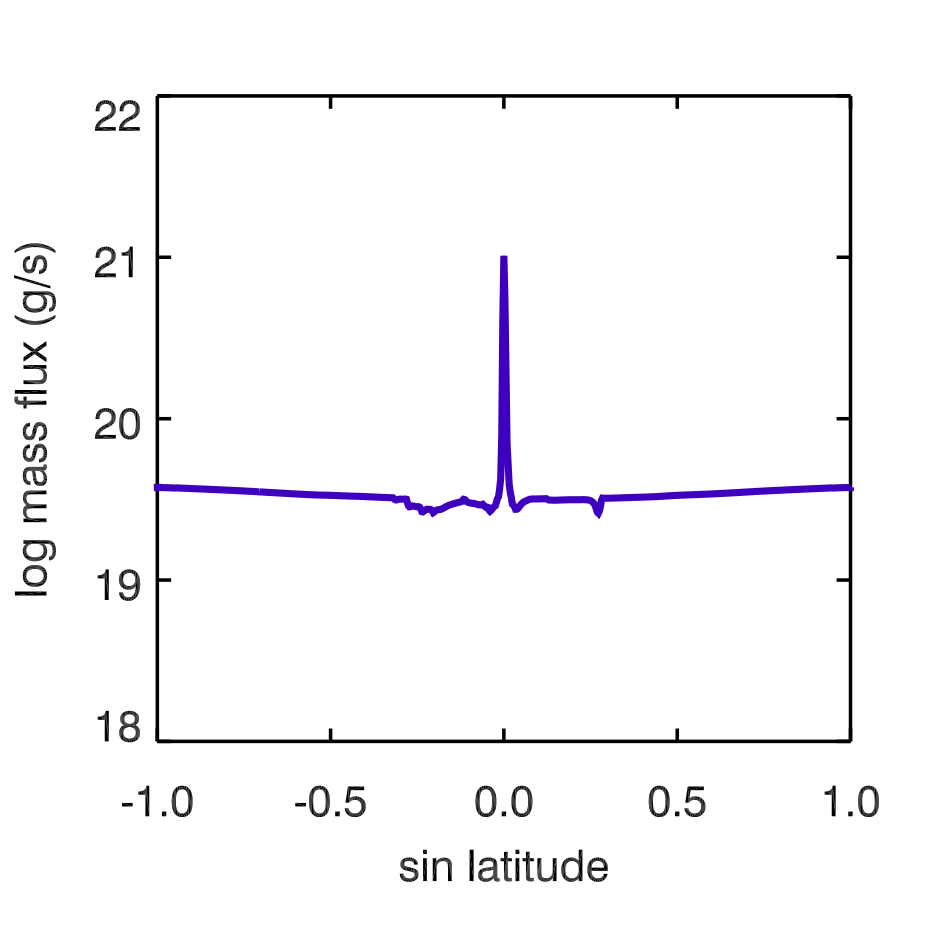}} \\
 \mbox{ \includegraphics[width=50mm]{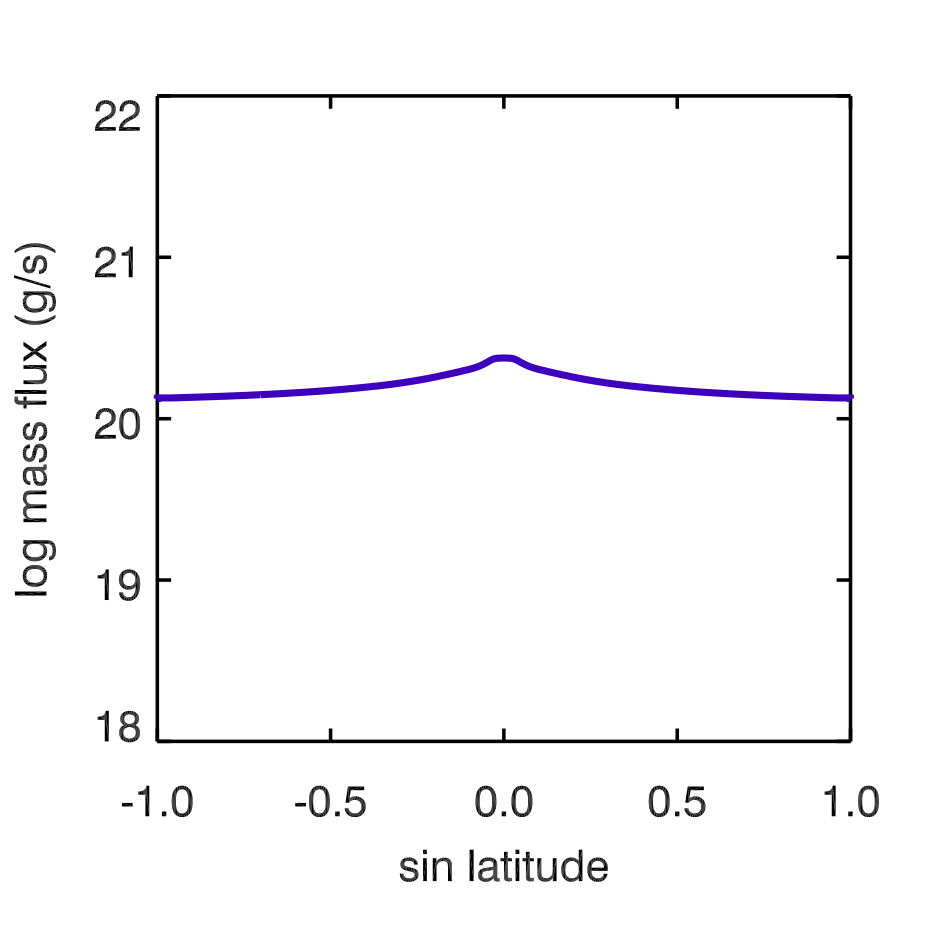} 
  \includegraphics[width=50mm]{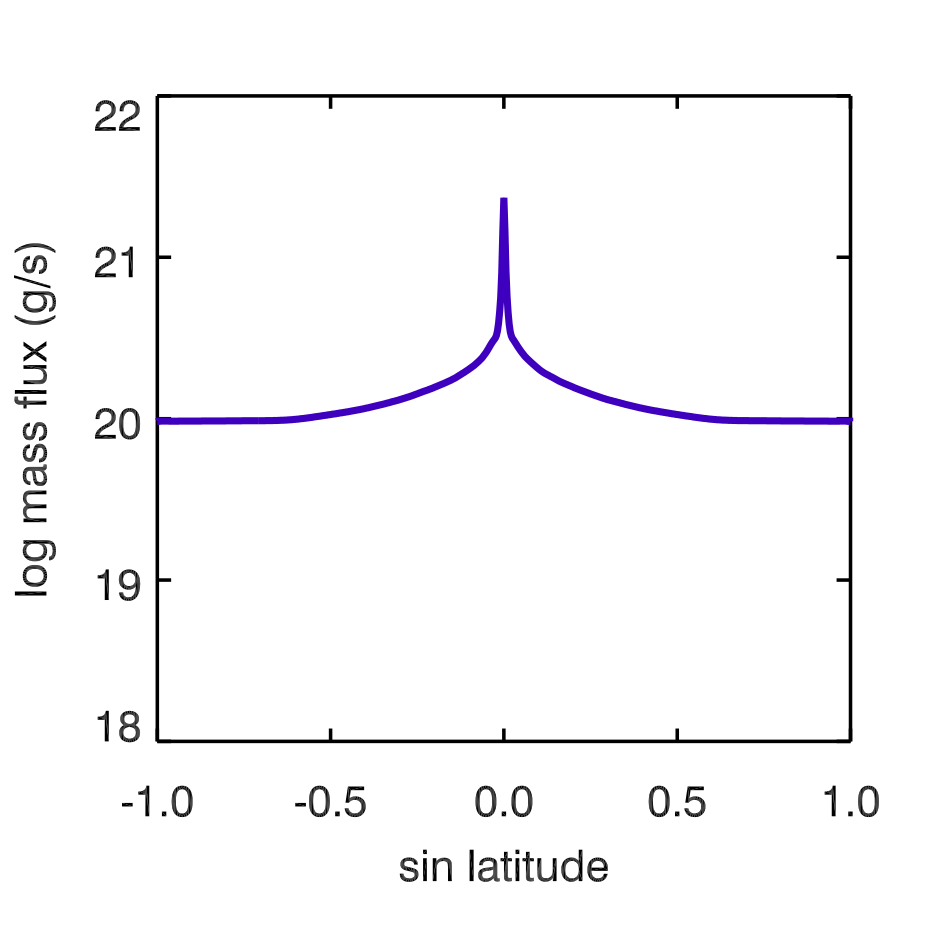} 
  \includegraphics[width=50mm]{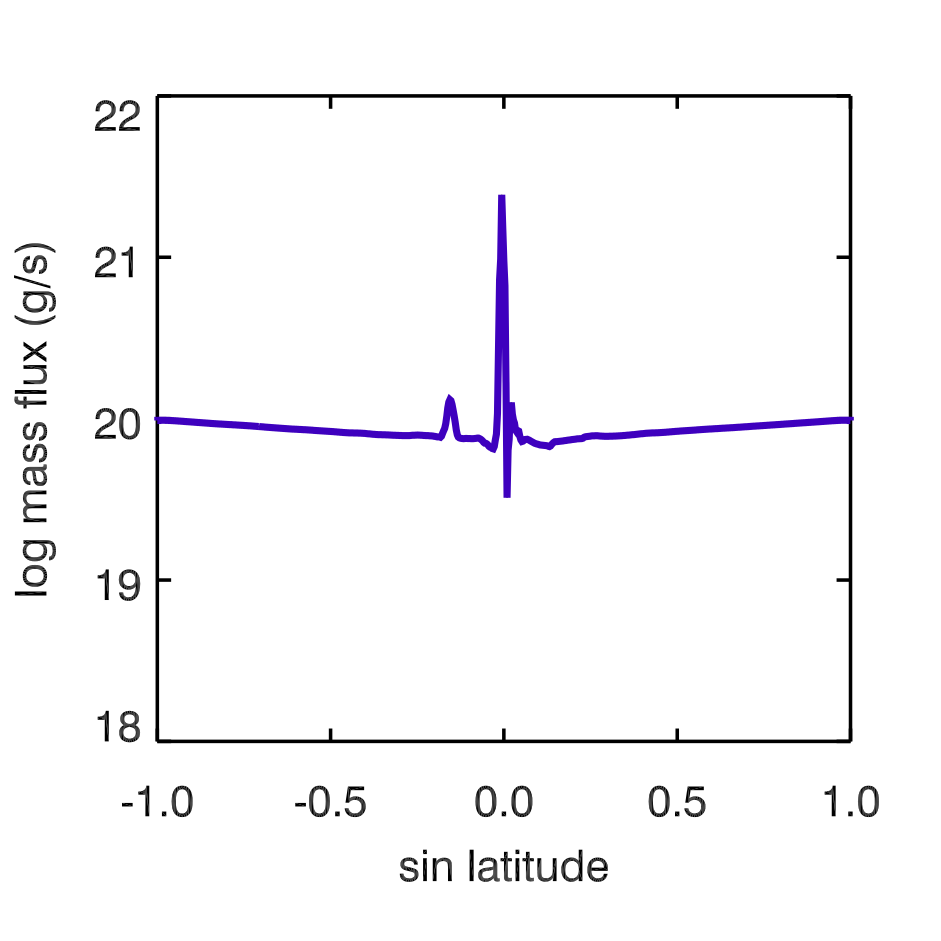}}
 \end{center}
 \caption{Radially-averaged mass flux as function of (sine) latitude for WR3 (top) and WR105 (bottom). Left $\eta=0.1$. Centre:  $\eta=1$. Right $\eta=10$.
   }
   \label{wr_mflux}
   \end{figure*}
Figure \ref{wr_mflux} shows the radially-averaged mass fluxes for WR3 and WR105.  As for $\zeta$ Pup, there is  a slight increase in the equatorial plane for weak confinement. For $\eta_*=1$  there is a sharp peak in the equatorial plane. Away from the equatorial plane the mass flux falls off with latitude, i.e.~is lowest at the poles.  
 For $\eta_*=10$ the peak is a bit more pronounced. For WR3 the mass flux away from the peak is now flat at low latitude while the falloff is still present at high latitude. WR105 shows a more pronounced peak than WR3, a somewhat jagged profile at low latitudes and an increase with latitude towards the poles. 
 
As the disk is very thin, the total mass loss is dominated by the wind despite the much larger values of the mass flux in the disk. For WR105, the fracture of the mass loss going through the region from +5 to -5 degrees of latitude, which accounts for 8.7 percent of the surface at a given radius, is 12.7 percent for  $\eta_*=0.1$. For $\eta_*=1$ this fraction increases to 25.6 percent. For $\eta_*=10,$ 23.7 percent of the mass loss go through the equatorial region. For WR3, the corresponding values are 13.3, 22.7, and 13.5 percent, respectively, for $\eta_*=$ 0.1, 1, and 10. In this case the mass flux is actually more concentrated at low latitudes for intermediate field strength than for the strongly confined case.   
 
\begin{figure}
 \begin{center}
  \mbox{
    \includegraphics[width=70mm]{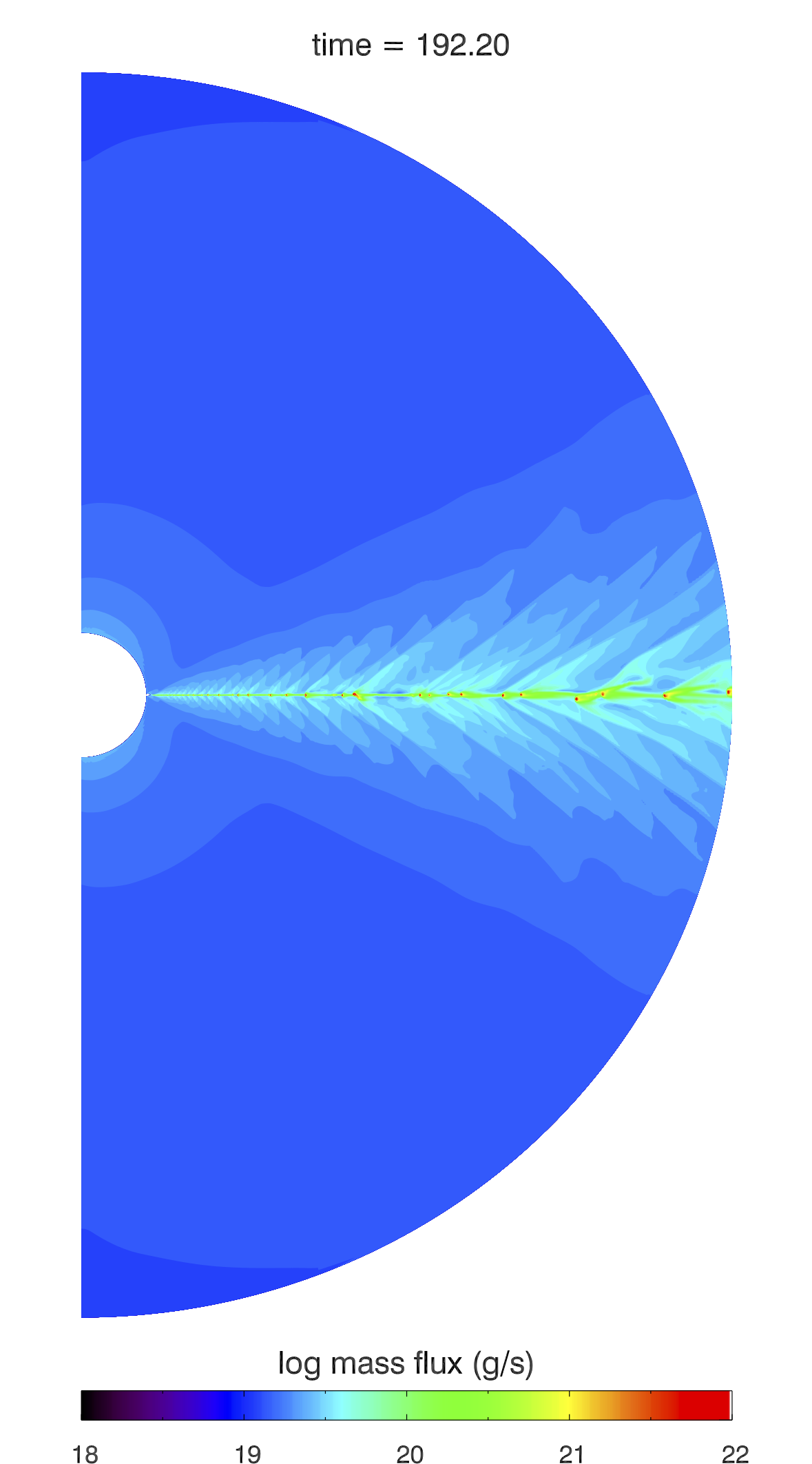}
 }
 \end{center}
 \caption{Radial mass flux for the  WR3  model with $\eta_*$=1.
 }
 \label{wr3_total}
 \end{figure}

This is further illustrated by Fig.~\ref{wr3_total}, which shows the radial mass flux distribution in the simulation box for $\eta_*=1$. 
While in the $\eta_*=0.1$ case there is already a pronounced concentration towards the equatorial plane, Fig.~\ref{wr3_total} shows increased
mass flux  at low latitudes, but now there is also the disk in the equatorial plane as a region of enhanced mass loss. The disk is fragmented and 
areas of increased mass density a the sources of a wake pattern that is also visible in the mass density.
For $\eta_*=10$, there is no wake pattern, the disk appears thinner, and there is little variation above and below the disk. In all cases mass loss in the equatorial plane in enhanced as the higher density outweighs the slower outwards motion. As the blobs in the disk plane form sporadically and move outwards, the mass flux in the disk (and thus the height of the peaks in Fig.~\ref{wr_mflux} varies with time. With increasing field strength, the total mass loss rate also becomes more variable as more and more gas is trapped in the closed field line region and is only released when the field configuration close to the star changes. Note that for $\eta_*=10$, the gas flow is mostly inwards near the inner boundary. The plots only show the outwards flows. 
\section{Discussion}   
\begin{figure}
\begin{center}
\includegraphics[width=8.0cm]{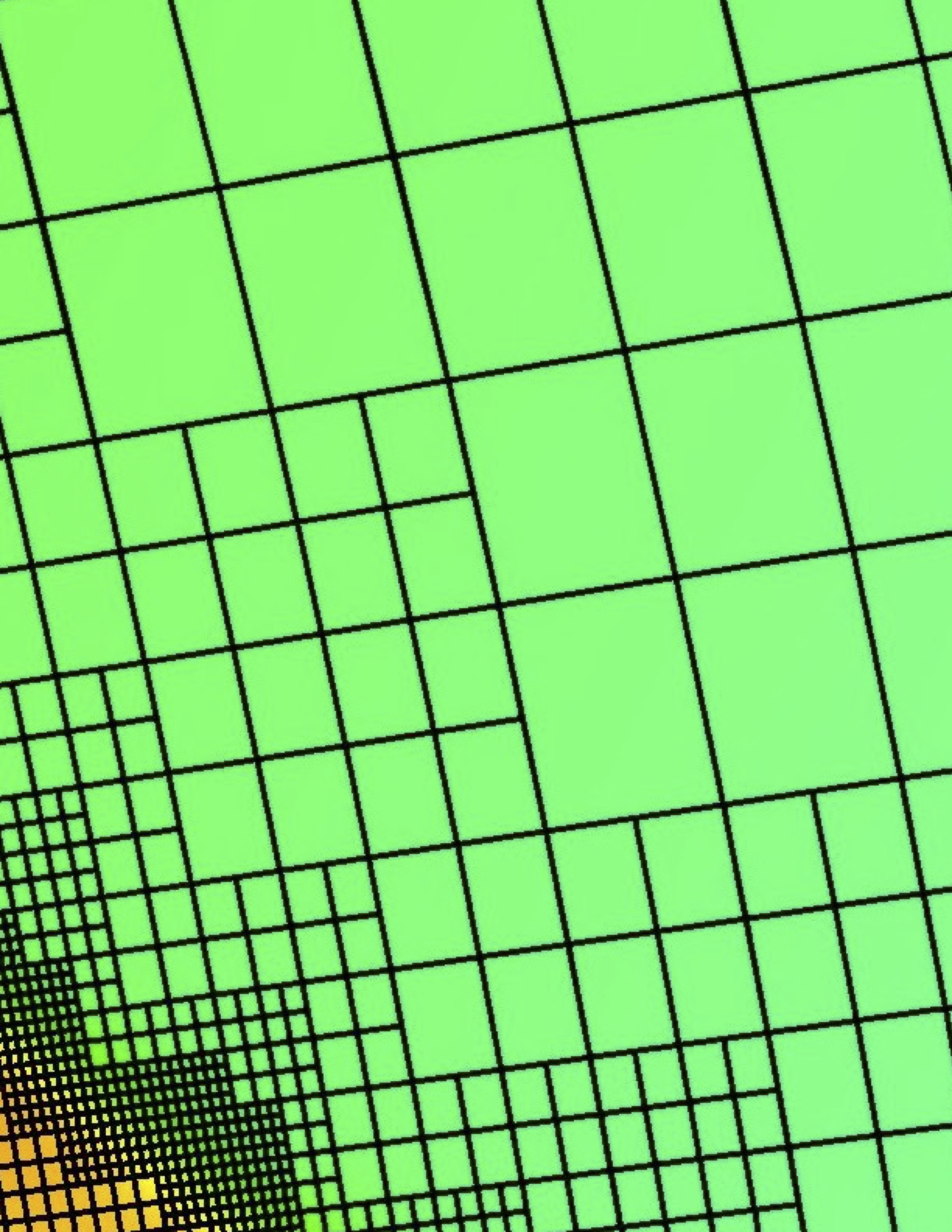}
 \end{center}
 \caption{Detail from a snapshot of the density stratification with the mesh overlaid.}
 \label{AMR}
\end{figure}

We have applied our setup to a number of massive stars spanning a range of effective temperatures from 23,000 K to 89,000 K, a range of luminosities from $3.57 \times 10^3 L_\odot$ to $10^6 L_\odot$, and mass loss rates from $1.5\times 10^{-10} M_\odot/a$ to $5 \times 10^{-5} M_\odot/a$. 
Our results confirm the concept of the confinement parameter, $\eta_*$ insofar as we find very similar magnetic field configurations and flow patterns
for very different objects provided the values of the confinement parameter is similar. 
In the test case of \zetpup{}, we have largely reproduced the findings of \cite{udd02}, who also reported a flattening of the density contours in case of weak fields and a steepening accompanied with the formation of a disc for stronger magnetic fields. 
 Fragmetation is more pronounced in the equatorial disk, though, and a wake-like pattern in the density distribution above and below the disk. 
 The reason for these differences is not quite clear, but the different layouts of the mesh may play a role. While the Zeus code allows a non-equidistant
 mesh in order to resolve the equatorial plane and the region above the stellar surface, the Nirvana code requires the basic mesh to be equidistant in both dimensions. Resolution of boundary layers or areas where the variables change fast over small distances, like the equatorial disk, is increased as needed by the adaptive mesh refinement algorithm. The mesh is highly refined near the inner boundary, in the equatorial plane and in the wake  pattern originating from the blobs in the disk plane.  Figure \ref{AMR} shows a detail from a snapshot of the mass density distribution with the mesh overlaid. The largest cells represent the basic mesh while the smaller cells are the result of adaptive mesh refinement. With five refinement levels, there is total of six cell sizes.
 
  Differences could also be caused by general differences in the properties of the numerical schemes used, like numerical diffusion, the implementation of the line force, or the treatment of the boundaries. Our results for the weak field case differ insofar from earlier work as we find a disk for the $\eta_*=0.1$ case. However, that model needed a run time of $10^4$ ks for the disk to form.

The results for the Wolf-Rayet stars are qualitatively similar to each other and those previously found for $\zeta$ Pup with the same values of the confinement parameter, $\eta_*$, thus confirming that the impact of the magnetic field on the wind indeed scales with this parameter.
At this point there is no Wolf-Rayet star with an unambiguously detected magnetic field.  \cite{chevrotiere13}  studied the WN4 star EZ XMa = WR6 = HD 50896). 
Table \ref{parameters} shows that very large field strengths would be needed to reach strong confinement in some cases, e.g.~WR 6. 
On the other hand, there are cases like WR105 where field strenghts of the order1 kG would be sufficient for strong confinement.  

Using the CAK line force for the Wolf-Rayet may stretch the boundaries of the validity of that model. Given the current observational uncertainties, however,  that formulation of the line force is to be seen more as a means to reproduce the observed wind properties rather than a correct physical description of the wind driving mechanism. A more realistic description should be adopted in future studies, though. 

\subsection{Effect of mean molecular weight and stellar mass}
\begin{figure}
 \begin{center}
  \includegraphics[width=80mm]{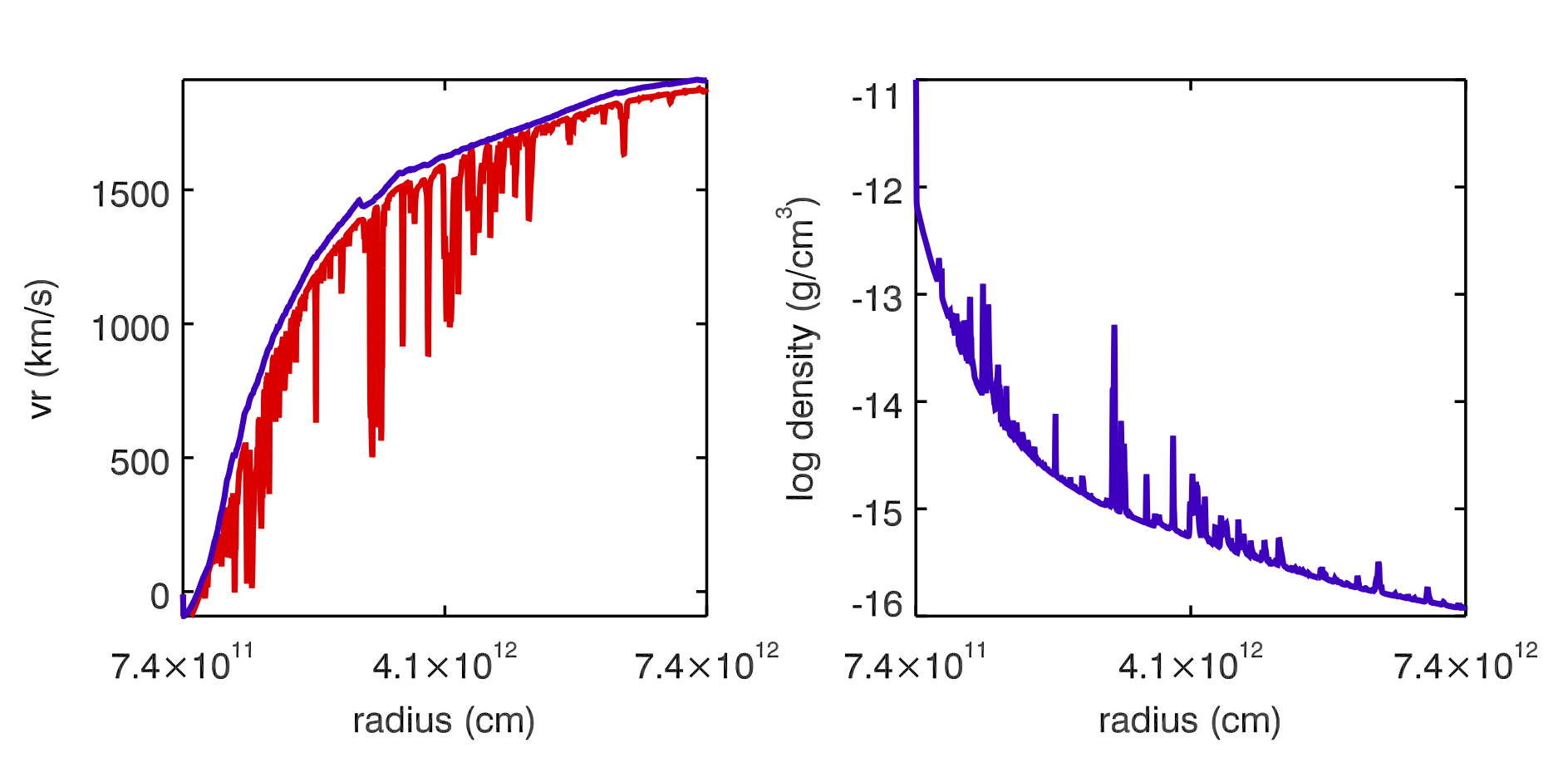} \\
   \includegraphics[width=80mm]{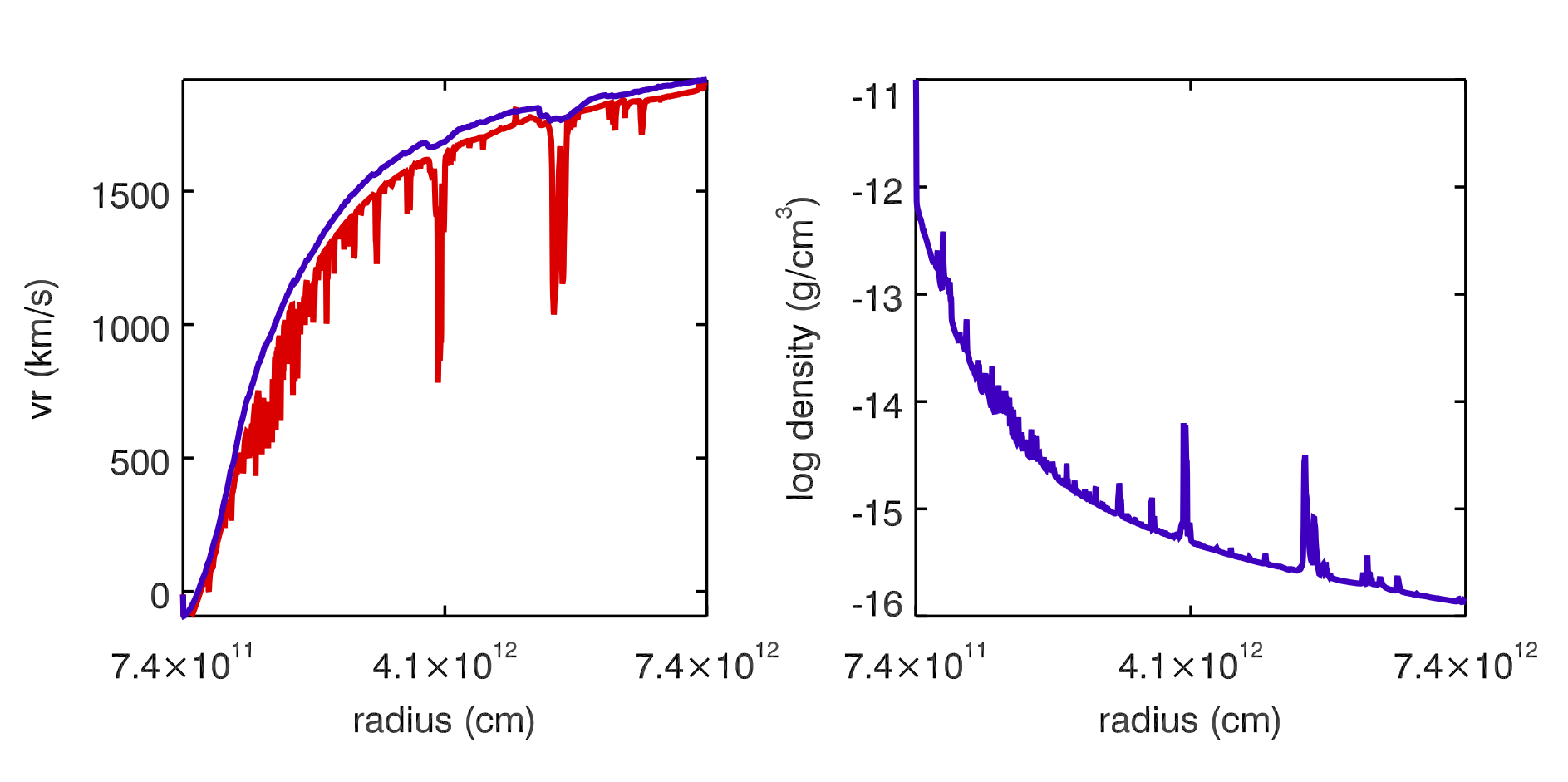} \\
   \includegraphics[width=80mm]{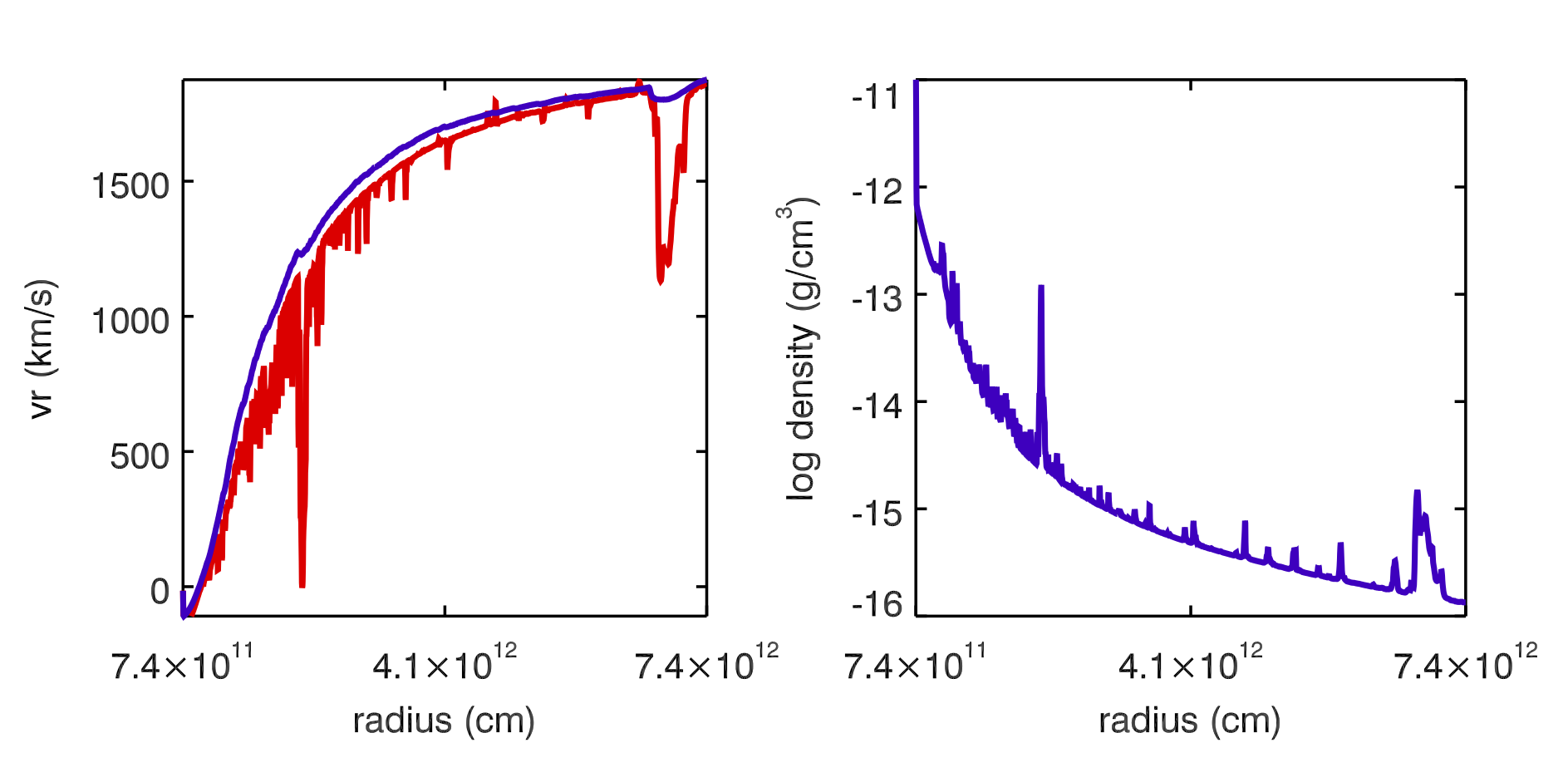}
 \end{center}
 \caption{Latitudinally-averaged radial velocity (left) and gas density (right) as functions of radius for \tetori{}. Top: $\mu=0.62$, middle: $\mu=1$, bottom: $m=45$\msol{}, $\mu=1$. The red lines mark the latitudinally-averaged radial velocity with the gas density used as a weight function.}
 \label{mucomp_wind}
\end{figure}

%
Throughout this paper we have so far used  a value of one for the mean molecular weight $\mu$. This obviously is not the correct value for winds originating from main sequence stars like the Sun, where you expect values around 0.6 but values closer to one are appropriate for evolved stars. Instead of assigning an individual value to each star, we have computed an alternative model for $\theta^1$ Ori C with $\mu = 0.62.$ 
 The falloff of the density with increasing distance from the stellar surface is very similar but the closed field line region is slightly larger for the $\mu=1$ model. Figure \ref{mucomp_wind} shows the latitudinally averaged  radial velocity and density vs.~radius as in Fig.~\ref{zetpup_winds}. The $\mu=0.62$ and $\mu=1$ cases are essentially identical, the only difference being the position of the clumps in the disk. Both models were calibrated to produce $v_\infty=3325$ km/s. With $\mu=0.62$, this required $\bar{Q}=560$ instead of 850. We conclude that the exact value of $\mu$ is of minor value in the type of model discussed here. For given temperature and density, a larger value of mu corresponds to a lower gas pressure, which seems to be the reason for the larger closed field line region in the $\mu=1$ case.

Likewise, the stellar mass can be varied to some degree without changing the wind pattern, provided the $\alpha$ and $\bar{Q}$ parameters are again adjusted to meed the observed terminal speed and mass loss rates. 
The  bottom panel in Fig.~\ref{mucomp_wind} shows a model with a stellar mass of 45\msol{} instead of 33.5\msol{}. 
With $\alpha=0.46$ and $\bar{Q}=635$, the wind structure is quite similar to that with the less massive star. As the stellar mass does not enter the confinement parameter, the surface field strengths are identical. 
 %

 \subsection{Split monopole geometry? }
Far away from the star the kinetic energy dominates over the magnetic field energy. The magnetic field therefore follows the gas motion and is directed radially outwards (or inwards, depending on the latitude) from the star.  This leads to a field configuration that is profoundly different from the initial dipole field. 
If one assumes a spherically symmetric wind, the field configuration closest to the symmetry of the outflow
is the split monopole, i.e.
\begin{equation}
  \vec{B} =  \pm \frac{B_0}{r^2} \vec{e_r}, 
\end{equation}
with the sign changing in the equatorial plane, as used by  \cite{gayley10} to compute the circularly polarized Stokes V profile for emission lines in hot star winds with weak radial magnetic field.

\begin{figure*}
 \begin{center}
  \mbox{
   \includegraphics[width=55mm]{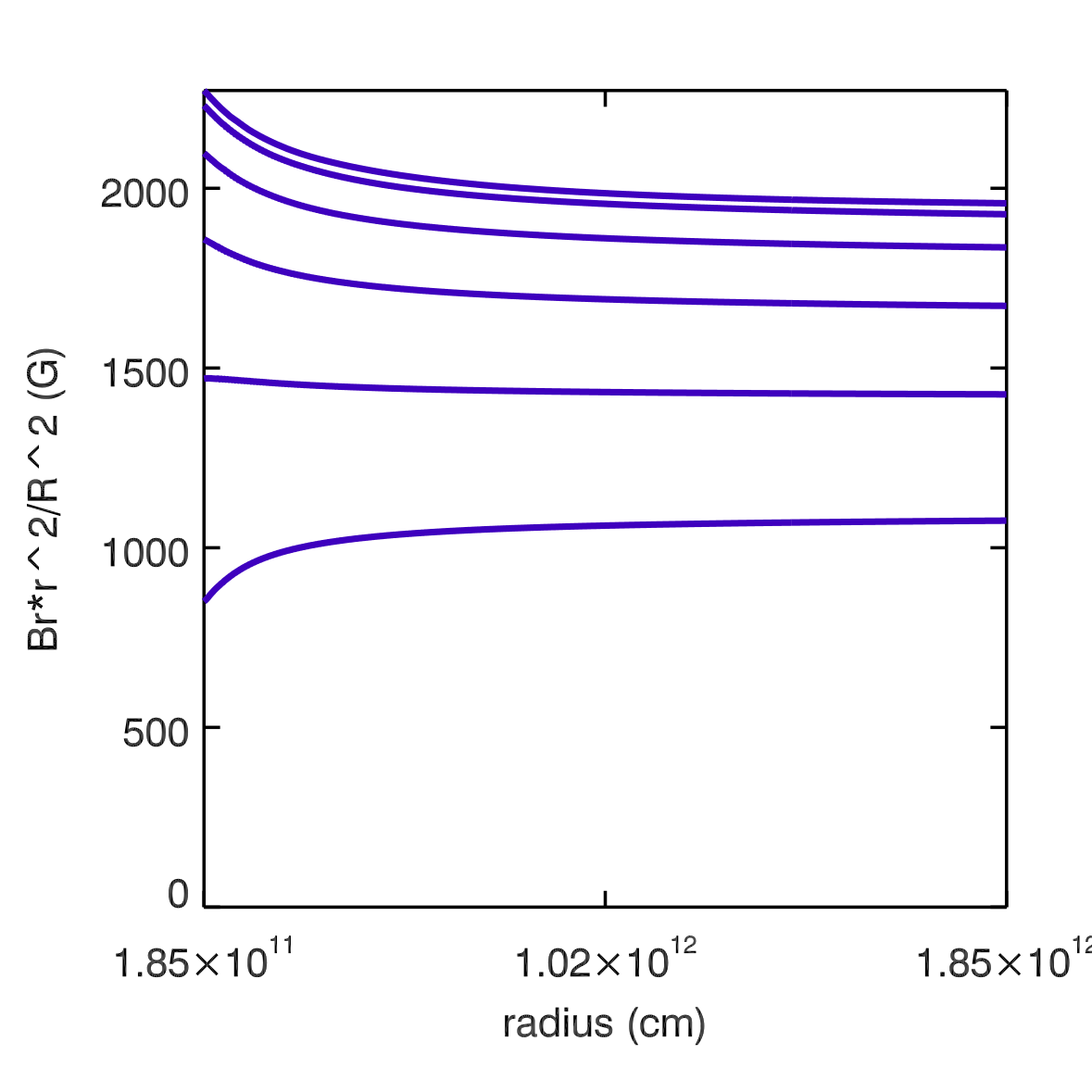} 
   \includegraphics[width=55mm]{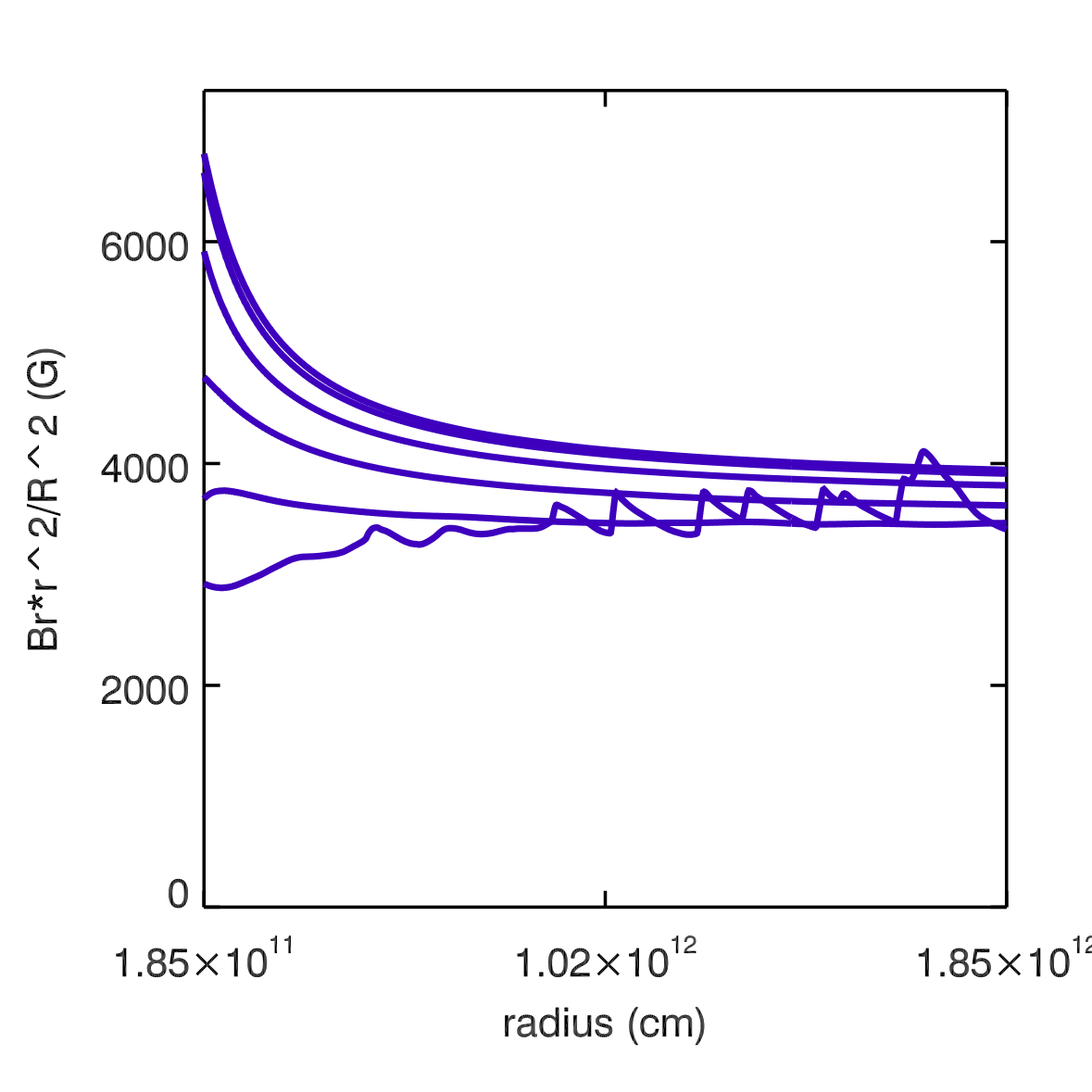} 
    \includegraphics[width=55mm]{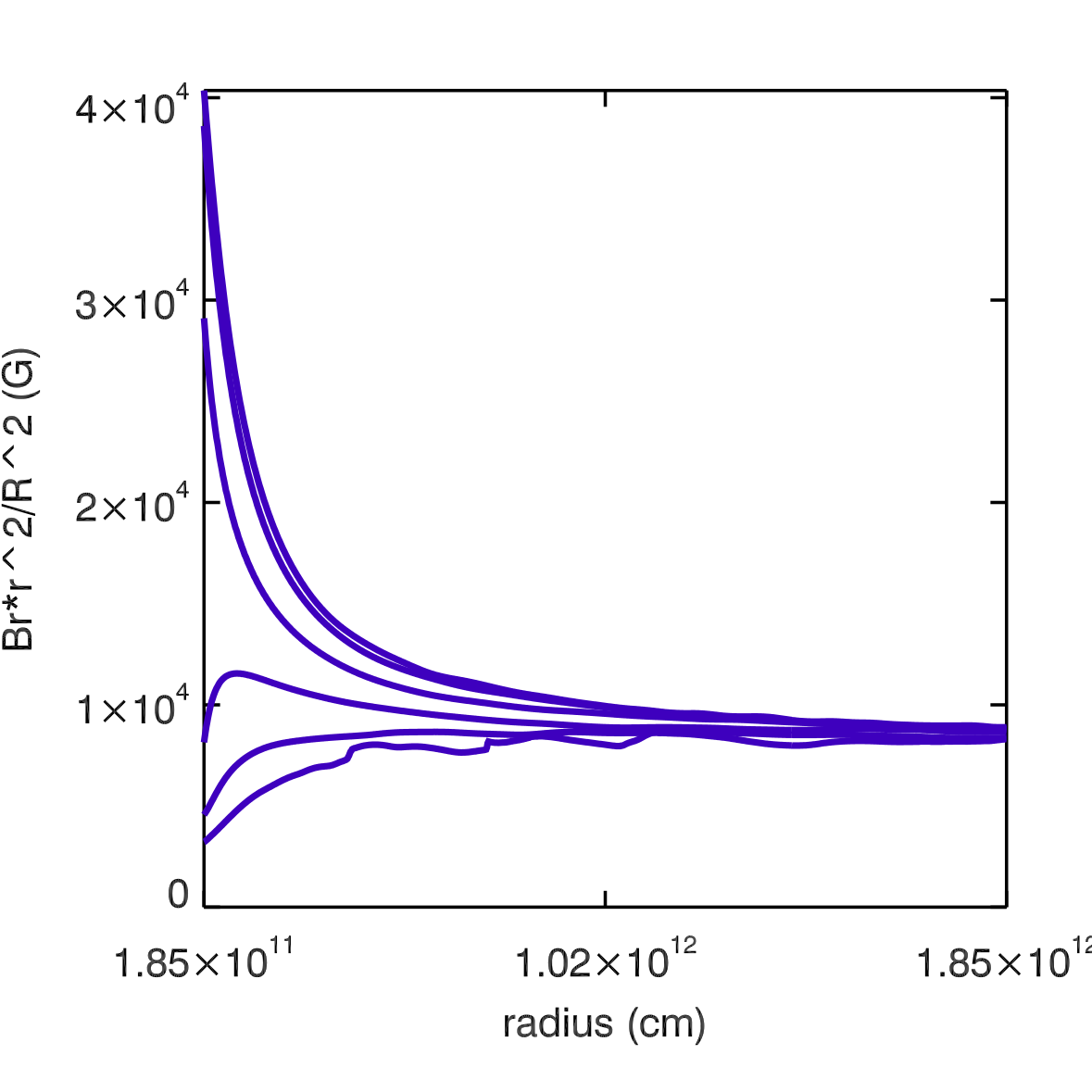}
    }
 \end{center}
 \caption{Radial component of the magnetic field multiplied by $r^2/R_*^2$ at 90, 75, 60, 45, 30, and 15 degree latitude (top to bottom) for WR6 with $\eta_*=0.1$ (left), $\eta_*=1$ (center), and $\eta_*=10$ (right).}
 \label{wr6_monopole}
\end{figure*}

As the magnetic field is divergence-free,  a purely radial field has to fall off with $r^2$, like a monopole field. 
However, an abrupt sign change in the equatorial plane requires a current sheet there. 
Figure \ref{wr6_monopole} shows $r^2 B_r $ vs. r at various  latitudes for the weak, intermediate, and strong confinement cases of the WR6 model. 
Far away from the star the lines are indeed horizontal, indicating that the falloff with radius follows the $1/r^2$ law. 
However, only the case with strong confinement shows a true split monopole geometry insofar as the magnetic field strength does not depend on latitude except for the sharp transition in the equatorial plane. 
In the weak confinement case the magnetic field preserves almost the full latitude dependence of the dipolar field at the stellar surface even at large radii. In the intermediate case the latitude-dependence at large radii is much reduced and the sign changes abruptly in the disk plane. Note that the line representing 15 degree latitude shows a sawtooth pattern, indicating a more complex field geometry at low latitudes. 
Note also the sharp decline of the magnetic field close to the star at high latitudes in the strong confinement case caused by the more dipolar field structure and the existence of closed loops there.

As the right panel in Fig.~\ref{wr6_monopole} shows, the field strength at large radii represents the surface field at 30 degree latitude rather than the pole, with a strength (of the radial field component) that is only 1/4th of the polar field strength. This could mean that surface fields of WR are stronger than implied from spectroscopy on the optically thin part of the wind assuming split monopole geomtery  \citep{chevrotiere13, chevrotiere14}.  
\section{Conclusions}
We have presented a numerical model for the mass loss of magnetic massive stars and shown that it largely agrees with previous work when the same assumptions are made. The confinement parameter \confin{} defined by \cite{udd02} proves to be a very reliable predictor of the way the magnetic field changes the stellar wind over a range of stellar and wind parameters. A weak dipolar field, as characterised by $\eta_*\ll1$ will be stretched out and become a split monopole, with a current layer in the equatorial plane. The mass flow remains largely spherically symmetric, but a thin disk forms in the equatorial plane. For intermediate field strength, $\eta_* \approx 1$, the equatorial disk is more prominent but fragmented while the magnetic field is still open. For strong magnetic fields, $\eta_*\gg1$, the magnetic field remains closed at low latitudes and gas will be trapped in this region of closed field lines. Mass loss continues largely unhindered at high latitudes. The volume of closed field lines turns out to be much smaller than that enclosed by the Alf\'en surface, $\eta=1$. Consequently, we find that the overall mass loss rate is only moderately reduced. 

The model presented here contains a number of simplifications, most notably axisymmetry, lack of rotation, and isothermy. Future work will have to drop these simplifications. Finally, the time-dependent and non-axisymmetric character of the solutions as well as the presence of inward flows call for a more sophisticated implementation of the radiative force than the CAK formalism we have used here.
\bibliography{magwind}

\end{document}